\documentclass[12pt]{article}
\usepackage{graphicx}
\usepackage{subfigure}
\usepackage[dvips]{color}

\input epsf.tex

\newcommand{\beq}{\begin{equation}}
\newcommand{\eeq}{\end{equation}}
\newcommand{\be}{\begin{equation}}
\newcommand{\ee}{\end{equation}}
\newcommand{\bea}{\begin{eqnarray}}
\newcommand{\eea}{\end{eqnarray}}

\makeatletter
\renewcommand{\theequation}{\thesection.\arabic{equation}}
\@addtoreset{equation}{section}
\makeatother

\def\href#1#2{#2}

\begin{document}

\baselineskip=15.5pt
\pagestyle{plain}
\setcounter{page}{1}

\begin{titlepage}
\begin{flushleft}
       \hfill                       FIT HE - 11-03 \\
       \hfill                       KYUSHU-HET 132 \\
\end{flushleft}

\begin{center}
  {\huge Holographic Glueballs and Infrared Wall   \\ 
   \vspace*{2mm}
Driven by Dilaton  \vspace*{2mm}
}
\end{center}

\begin{center}

\vspace*{2mm}
{\large Kazuo Ghoroku${}^{\dagger}$\footnote[1]{\tt gouroku@dontaku.fit.ac.jp},
Kouki Kubo${}^{\ddagger}$\footnote[2]{\tt kkubo@higgs.phys.kyushu-u.ac.jp},
Tomoki Taminato${}^{\ddagger}$\footnote[3]{\tt taminato@higgs.phys.kyushu-u.ac.jp},
\\
and ${}^{\P}$Fumihiko Toyoda\footnote[4]{\tt ftoyoda@fuk.kindai.ac.jp}
%
}\\

\vspace*{2mm}
{${}^{\dagger}$Fukuoka Institute of Technology, Wajiro, 
Higashi-ku} \\
{
Fukuoka 811-0295, Japan\\}
{
${}^{\ddagger}$Department of Physics, Kyushu University, Hakozaki,
Higashi-ku}\\
{
Fukuoka 812-8581, Japan\\}
{
${}^{\P}$Faculty of Humanity-Oriented Science and
Engineering, Kinki University,\\ Iizuka 820-8555, Japan}

\vspace*{3mm}
\end{center}

\begin{center}
{\large Abstract}
\end{center}
We study glueballs in the holographic gauge theories,
supersymmetric and non-super symmetric cases, which are given by
the type IIB superstring solutions with non-trivial dilaton. In both cases, the dilaton 
is responsible for the linear 
potential between the quark and anti-quark, then we could see the meson spectra.
On the other hand, the glueball spectra are found for the non-supersymmetric case, but
not for the supersymmetric case.
We find that we
need a sharp wall, which corresponds to an infrared cutoff, in order to obtain the glueballs.
In the non-supersymmetric case, the quantized glueballs are actually
observed due to the existence of such a wall driven by the dilaton. 
We could see the Regge behavior of the higher spin glueball states, and
the slope of the glueball trajectory is half of the flavor 
meson's one.

\noindent

\vfill
\begin{flushleft}

\end{flushleft}
\end{titlepage}
\newpage

\vspace{1cm}
\section{Introduction}
Since the 
holographic approach is a
powerful method to study the non-perturbative properties of 
the strong coupling gauge theories \cite{ads1,ads2,ads3},
various attempts to examine the properties of quantum chromo
dynamics (QCD) have been performed. Among them, an interesting approach
is to study 
the mass spectra of glueballs. The discrete mass spectra for small spin states
are obtained in terms of the normalizable
Kaluza-Klein modes of quantum fluctuations of bulk fields \cite{ORT}-\cite{Gursoy:2007er}.

Previously we have studied the open strings (flavored mesons)  \cite{GTT}
in the background with two simple dilaton configurations, supersymmetric  \cite{KS2,LT}
(SUSY) and non-supersymmetric (non-SUSY) \cite{background} versions.
These configurations
are obtained as the solutions of type IIB supergravity with five form field flux, 
dilaton (and axion for SUSY version). In both solutions, 
the non-trivial dilaton provides the gauge condensate $\langle F^2\rangle$, and it leads to
the tension ($\tau_M$) of the linear potential
between the quark and anti-quark being proportional to $\sqrt{\langle F^2\rangle}$ \cite{GY}.
Then we could obtain the mass spectra for 
the flavor mesons. Further, the Regge behavior for 
their higher spin states has been obtained in terms of 
rotating open-string configurations for the Nambu-Goto action \cite{GTT}.

\vspace{.3cm}
Here, the analysis is extended to the glueballs (closed strings) 
in the two dilatonic backgrounds mentioned above in order to make clear the role
of the dilaton or the gauge condensate furthermore. 
We firstly show the existence of a kind of potential wall
for strings in the non-SUSY bulk, but it is absent in the SUSY case. 
Then we find the importance of
this wall to realize the glueball spectra through the following two analyses.

Firstly, 
the classical configuration is studied for the closed spinning strings
as performed in other cases \cite{KMMW,ZSV,BCMZ,KZSV,PT,KV,Huang:2007fv}.
Supposing the folded configuration, we find the Regge behavior 
for both SUSY and non-SUSY. 
However, in the SUSY case, the stable configuration is found at
the horizon ($r=0$) where a metric singularity is observed for $g_{rr}(r)$ with respect to the 
radial coordinate $r$. 
This singularity implies that the quantum fluctuation of this
configuration in the direction of $r$ is suppressed, then we expect the absence of the glueball 
obtained from the bulk fields.

This point is assured through the second glueball analysis, in which 
the glueballs are studied as the quantum fluctuations of bulk fields.
Namely, while the glueballs are observed in the non-SUSY
case, but not in the SUSY case for various bulk fields.
Both analises are therefore compatible.
As a result, it can be said that the SUSY background considered here is not enough to realize the glueballs in spite of the fact that the quark confinement is realized in this case.

\vspace{.3cm}
In the non-SUSY background, however, the bulk curvature 
has a (naked) singularity at $r=r_0$. Then, we should perform the holographic
analysis in the region of $r$ outside of this singularity. 
In the case of D4 brane model \cite{ORT,Minahan}, the singularity (at $r=0$)
is covered by the event horizon $r_h$, and the holographic analysis is restricted
to $r\geq r_h(>0)$.
In the non-SUSY background considered here, 
such a parameter like $r_h$ is absent in the metric, 
however, we observe that 
strings and branes are prevented to arrive at $r_0$ due to the wall mentioned above.
For example, we find that static open-strings are blocked at $r_m(>r_0)$, and then
the rotating closed strings are trapped at $r_m$, where the metric
is non-singular.

Another type of non-SUSY background solution, which also has a naked singularity at $r=0$,
has been proposed by Constable and Myers \cite{CoM}, and then it has been used to study the
meson spectrum \cite{Babington:2003vm}.
In this case also,
a similar potential wall can be observed, and then
This wall prevents any physical fields
in the bulk from approaching to the naked singularity. However, in this case, 
glueballs coming from the graviton and dilaton were absent
\cite{CoM} in spite of the wall. Then, in this point, the dual theory of \cite{CoM}
is different from our non-SUSY case
since glueballs for all fluctuations are observed in our case. 


\vspace{.5cm}

The outline of this paper is as follows.
In the next section, the bulk solutions for our holographic model are given, then the wall
of the gravitational potential is shown and examined for strings.
In the section 3, glueballs are studied by solving the Nambu-Goto action, and then
the role of the potential wall is shown. Then, in the next section, 
the glueball spectra are given by solving the equations of motion of
the bulk field fluctuations. The results are compared with the lattice simulation and 
other calculations in the section 5.
The summary and discussions are given in the final section.

\section{Bulk Background}\label{sec:Bulk_Background}

Here we give the ingredient of our holographic model for confining
Yang-Mills theory to study glueballs.
We consider 10D IIB model retaining the dilaton
$\Phi$, axion $\chi$ and self-dual five form field strength $F_{(5)}$.
{The action is given as
\beq\label{2Baction}
 S_{(10)}={1\over 2\kappa^2}\int d^{10}x\sqrt{-g}\left(R-
{1\over 2}(\partial \Phi)^2+{1\over 2}e^{2\Phi}(\partial \chi)^2
-{1\over 4\cdot 5!}F_{(5)}^2
\right), \label{10d-action}
\eeq
where other fields are consistently set to zero,  and 
$\chi$ is Wick rotated \cite{GGP}.}
Under the Freund-Rubin
ansatz for $F_{(5)}$, 
$F_{\mu_1\cdots\mu_5}=-\sqrt{\Lambda}/2~\epsilon_{\mu_1\cdots\mu_5}$ ,
and for the 10D metric as $M_5\times S^5$ or
$ds^2=g_{\mu\nu}dx^{\mu}dx^{\nu}+g_{kl}dx^kdx^l$, the solution given below
has been found \cite{KS2,LT}.
Where $(\mu,\nu)= 0\sim4$ and  $(k,l)=5\sim9$.
The five dimensional part ($M_5$) of the
solution is obtained by solving the following reduced 5D action,
\beq
 S_{(5)}={1\over 2\kappa^2}\int d^5x\sqrt{-g}\left(R+3\Lambda-
{1\over 2}(\partial \Phi)^2+{1\over 2}e^{2\Phi}(\partial \chi)^2
\right), \label{5d-action}
\eeq
which is written 
in the Einstein frame. 
{And the corresponding equations of motion are given as
\beq\label{metric5d}
 R_{MN}={1\over 2}\left(\partial_M\Phi\partial_N\Phi-
         e^{2\Phi}\partial_M\chi\partial_N\chi\right)-\Lambda g_{MN}
\eeq
\beq \label{p5d}
 {1\over \sqrt{-g}}\partial_M\left(\sqrt{-g}g^{MN}\partial_N\Phi\right)=
    -e^{2\Phi}g^{MN}\partial_M\chi\partial_N\chi\ , 
\eeq 
\beq\label{chieq5d}
   \partial_M\left(\sqrt{-g}e^{2\Phi}g^{MN}\partial_N\chi\right)=0
\eeq}
The bulk solutions are obtained under the ansatz for the metric,
\bea 
ds^2_{10}&=&G_{MN}dX^{M}dX^{N}=e^{\Phi/2}g_{MN}dX^{M}dX^{N}\nonumber\\
&=&e^{\Phi/2}
\left\{
{r^2 \over R^2}A^2(r)\left(-dt^2+\sum_{i=1}^3(dx^i)^2\right)+
\frac{R^2}{r^2} dr^2+R^2 d\Omega_5^2 \right\} \ , 
\label{finite-c-sol}
\eea
where $G_{MN}$ ($g_{MN}$) denotes the string (Einstein) frame metric and $M,~N=0\sim 9$ and
$R=\sqrt{\Lambda}/2=(4 \pi g_s N_c{\alpha'}^2)^{1/4}=({\lambda{\alpha'}^2})^{1/4}$ and
$\lambda=4 \pi g_s N_c$ denotes the 'tHooft coupling.
We consider the following two simple solutions which are dual to
the confining YM theory.

\vspace{.3cm}
\noindent{\bf (i) Supersymmetric solution}

In order to reserve supersymmetry, the solution is obtained under the ansatz,
\beq
\chi=-e^{-\Phi}+\chi_0 \ .
\label{super}
\eeq
Then, we obtain
\beq\label{susy-solution}
e^\Phi= 1+\frac{q}{r^4}\ , \quad A=1\, ,
\eeq
where the dilaton is set as $e^{\Phi}=1$ at 
$r\to \infty$, and the parameter $q$ corresponds to the vacuum expectation value (VEV) 
of gauge fields strength~\cite{LT}
of the dual theory. Then this solution is dual to the four dimensional 
$\cal{N}$=4 SYM theory with a constant gauge condensate. Due to this condensate, the supersymmetry is reduced to $\cal{N}$=2 and then the conformal
invariance is lost since the dilaton is non-trivial as given above. As a result, 
the theory is in the quark confinement phase since
we find a linear rising potential between quark and anti-quark
with the tension $\sqrt{q}/(2\pi\alpha'R^2)$ \cite{KS2,LT,GY}.

Furthermore, we can see that the space-time is regular at any point.
In the ultraviolet limit, $r\to\infty$, 
the dilaton part $e^{\phi}$ approaches to one and 
the metric (\ref{finite-c-sol}) is reduced to $AdS_5\times S^5$. 
On the other hand, the dilaton part $e^{\phi}$ diverges in the infrared
limit $r\to 0$, so that one may expect a singularity at $r=0$. However there is
no such a singular behavior. 
This is assured by rewriting the metric (\ref{finite-c-sol}) 
in terms of new coordinate $z$, where $z=R^2/r$. Then we obtain 
\beq\label{background-2}
ds^2_{10}=e^{\Phi/2}{R^2\over z^2}
\left(-dt^2+(dx^i)^2+dz^2+z^2 d\Omega_5^2 \right) \ . 
\eeq 
In the infrared limit $z\to\infty$, we have
\beq\label{susy-metric-2}
  e^{\Phi/2}{R^2\over z^2}=R^2\sqrt{\frac{q}{R^8}+\frac{1}{z^4}}\sim \frac{\sqrt{q}}{R^2}\, . 
\eeq
Therefore we find 10D flat space time in this limit and no singular point
\cite{KS2,LT}.

\vspace{.3cm}
Here we notice the following point. The left-hand side of Eq.(\ref{susy-metric-2}) 
is expressed as
\beq
  e^{\Phi/2}{R^2\over z^2}=\sqrt{|G_{tt}|~G_{ii}}\equiv Q(z)\, ,
\eeq
where $G_{ii}$ expresses one of the metric of the three space component and is not
summed up. $Q(z)$
has a minimum at $z=\infty$ ($r=0$) and the value of the minimum is finite.
This is the condition to reproduce the area law of the Wilson loop \cite{KSS}.
The minimum value of $Q(z)$ is proportional to the tension of the linear potential
between the quark and the anti-quark. From this observation,
we can assure that the background given here leads to the confinement of the dual gauge theory.

\vspace{.3cm}
\noindent{\bf (ii) non-Supersymmetric solution}

As for the non-supersymmetric case, the solution is given 
by retaining only the dilaton, namely for $\chi=0$, then the supersymmetry 
is lost in this case. 
The solution is obtained as \cite{background}
\begin{equation}\label{non-susy-sol}
A(r)=\left(1-\left(\frac{r_0}{r}\right)^8\right)^{1/4},\qquad
e^{\Phi}=\left(\frac{(r/r_0)^4+1}{(r/r_0)^4-1}\right)^{\sqrt{3/2}},\qquad
\chi=0\, .
\end{equation}
This configuration leads to curvature singularity at the horizon $r=r_0$. 
So we cannot extend our analysis upto this horizon where some terms
like higher powers of curvatures or non-trivial RR fields
would be needed to make smooth the singularity. This point is
an open problem here.

\begin{figure}[htbp]
\vspace{.3cm}
\begin{center}
\includegraphics[width=12cm,height=8cm]{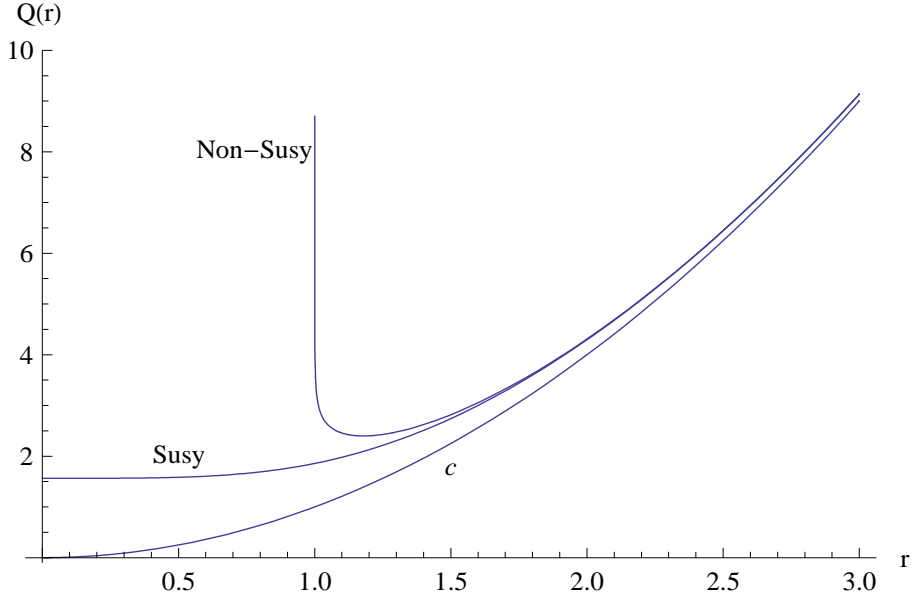}
\caption{{\small The gravitational potentials $R^2Q(r)$ for $r_0=1$ is shown
for non-supersymmetric solution, and for supersymmetric case with $q=\sqrt{6}r_0^4$. 
The minimum at about $r=1.18~r_0$ and the steep potential wall near $r_0$ are seen for non-supersymmetric case. The curve c ($=r^2$) represent
the potential for the case of $\Phi=$constant, namely for no gauge condensate.
}}
\label{g00-Pot}
\end{center}
\end{figure}

\vspace{.3cm}
The confinement property of the dual theory for this solution is 
assured as above through the
factor $Q=\sqrt{|G_{tt}|~G_{ii}}$. In this case, it is given as
\beq
  Q(r)=e^{\Phi/2}{r^2\over R^2}A^2(r)\mathop{=}_{r\to r_0}
           {2^{{1+\sqrt{3/2}}\over 2}\over \left(1-\left({r_0\over r}\right)^4\right)^{{-1+\sqrt{3/2}}\over 2}}\, .
\eeq
This diverges at $r=r_0$, then rapidly
decreases with increasing $r$ near $r_0$ (see Fig.\ref{g00-Pot}). On the other hand,
for large $r$, $e^{\Phi/2}$ and $A^2(r)$ approach to one, then 
$Q(r)$ increases with $r$ like $r^2$. These implies that
$Q(r)$ has a minimum at a point $r=r_m(>r_0)$.
Actually, from 
\beq
\frac{\partial Q(r)}{\partial r}{\Bigg|}_{r_m}=\frac{2e^{\Phi/2}}{R^2 r^7
A^2}\left(r^8-\sqrt{6}r_0^4r^4+r_0^8\right){\Bigg|}_{r_m}=0 ,
\eeq
$r_m$ and the minimum value $Q(r_m)$ are obtained as follows
\beq\label{rm}
 r_m=\left(\frac{\sqrt{6}+\sqrt{2}}{2}\right)^{1/4}r_0
\approx 1.18~r_0\, , \quad Q(r_m)\approx 2.40 \left({r_0\over R}\right)^2\, .
\eeq

\vspace{.3cm}
\noindent{\bf (iii) Wall for strings}

As shown below, the static open strings could approach to $r_m$, but they are blocked at $r_m$ since infinite energy
is needed to arrive at this point. 
In the following, we examine more details of this phenomenon.

The two-dimensional
world-sheet coordinates of open string are set as $(\tau,\sigma)=(t,x(r))$, and the
Nambu-Goto action for the open string is given as 
\bea
S&=&-\frac{1}{2\pi\alpha'}\int d\tau d\sigma \sqrt{-\det {\cal G}_{ab}}\\
&=&-\frac{1}{2\pi\alpha'}\int dt dx \sqrt{-G_{tt}\left(G_{xx}+G_{rr}(\partial_{x}r)^2\right)}\ ,
\eea
where ${\cal G}_{ab}=G_{MN}\partial_{a}X^M\partial_{b}X^N$.
Then, the energy is given as
\beq\label{general-wilson-E}
E=\frac{2}{2\pi\alpha'}\int_{0}^{L/2}dx\sqrt{-G_{tt}\left(G_{xx}+G_{rr}(\partial_{x}r)^2\right)}\ ,
\eeq
where $L$ represents the distance between the end point of the open
U-shaped string at the boundary $r=r_{max}$. The classical solution has been solved as
U-shaped form in the $x$-$r$ plane.
At the bottom of the U-shape string, $(x,r)=(0,r_{min})$, $\partial_{x}r=0$.
Therefore, $r=r_{min}$ is the minimum point
of the open string of $r$. 
Solving the equation of motion for $r$, we get the following relation
\beq
\frac{Q^2}{\sqrt{Q^2-G_{tt}G_{rr}(\partial_{x}r)^2}}=h, 
\eeq
where $Q^2$ is defined above and $h$ represents an integration constant. Then, this relation
becomes
\beq\label{sol-eom-wilson1}
\partial_{x}r=\pm\sqrt{H\left(\frac{Q^2}{h^2}-1\right)}\, , \quad
   H=\frac{G_{xx}}{G_{rr}}.
\eeq
Then we can take as
\beq
 Q(r_{min})=h\, .
\eeq
As a result, $L$ and $E$ are given as
\beq\label{wilson-length}
\frac{L}{2}=\int_{r_{min}}^{r_{max}}dr\frac{{1}}{\sqrt{H\left({Q^2\over h^2}-1\right)}}\ ,
\eeq
\beq\label{energy-sol}
E=\frac{1}{\pi\alpha'}\int_{0}^{L/2}dx {Q^2\over h}   \ .
\eeq

\vspace{.3cm}

For the background (\ref{non-susy-sol}), $L$ is estimated as
\beq\label{L-sep-int}
\frac{L}{2}=\left(\int_{r_{min}}^{r_{min}+\epsilon}+\int_{r_{min}+\epsilon}^{r_{max}}\right) {dr\over\sqrt{H\left({Q^2\over h^2}-1\right)}}\, ,
\eeq
where $\epsilon$ is a small finite number. The latter integral is finite since the integrand
is finite in the region of $(r_m<)r_{min}+\epsilon<r<r_{max}$. Then, the former
integral is estimated in the limit of $r_{min}\to r_m$ as follows
\footnote{The details of the calculation are given in Appendix \ref{sec:dsl}},
\bea\label{L-es}
  I_1 &\equiv & \lim_{r_{min}\to r_m}\int_{r_{min}}^{r_{min}+\epsilon}{dr\over\sqrt{H\left({Q^2\over h^2}-1\right)}} 
  = \left.\frac{Q}{\sqrt{\frac{1}{2}H\left(Q^{2}\right)^{\prime\prime}+H^{\prime}\left(Q^{2}\right)^{\prime}}}\right|_{r=r_{m}} \nonumber \\
 &\times &\left\{ \log\left({\epsilon}+p+\sqrt{{\epsilon}^{2}+2p{\epsilon}}\right)-
\log\left(p\right)\right\}\, , \nonumber  \\
\eea
where the right hand side is expanded near $r_{min}=r_m$, and
\beq 
p =
\left.\frac{1}{\left(Q^{2}\right)^{\prime\prime}/\left(Q^{2}\right)^{\prime}
+2H^{\prime}/H}\right|_{r=r_{m}}\ . \label{eq:b/a}
\eeq 
Here, for non-supersymmetric case, we notice that $\left.p\sim {\left(Q^{2}\right)^{\prime}\over \left(Q^{2}\right)^{\prime\prime}}\right|_{r=r_{m}}= 0$
and the prefactors is finite. Then we find
\beq
  \left. L\right|_{r\to r_{m}} \to \infty\, .
\eeq
Then the open strings cannot exceed $r_m$ due to the infinite energy cost since the
energy is proportional to the length of the string as shown below.

\vspace{.3cm}
As for the energy $E$, it would be estimated from (\ref{energy-sol}) at its minimum 
according to the action principle as follows,
\beq
  E\approx \frac{Q(r_{m})}{\pi\alpha'}\int_{0}^{L/2}dx
  =\frac{Q(r_{m})L}{2\pi\alpha'} \, .
\eeq
This implies that the tension of the linear potential between a quark and an anti-quark
is given for the non-supersymmetric case as
\beq
 \tau=\frac{Q(r_{m})}{2\pi\alpha'}\, .
\eeq
This result guarantees the confinement of quarks.

Here we give the following comments related to the above calculation\footnote{We also find more roles of the infrared wall for other classical
configurations, for instance, D7-brane embedding as flavor-brane and
D5-brane wrapped on $S^5$ as baryon vertex. Since its analyses are far
from our purpose, we give the details in Appendix \ref{more-on-wall}.}.

i) ~The same phenomenon
is seen for the supersymmetric case by replacing $r=r_m$ by $r=0$. In this sence, the wall
is receded to the limit of $r=0$, where $Q(r)$ defined above takes its finite
minimum value $Q(0)=\sqrt{q}/R^2$ due to the non-vanishing $q$. Then we could find
linear potential also in the supersymmetric case \cite{KS2,LT,GY}.

ii) ~In the case of the Witten's D$4$ model, we also find this behavior of the string stretching
at the event horizon. In this case, however, the origin of this behavior is reduced to the property $H=0$ 
at the blackhole horizon. In this sense, the mechanism of the confinement in D4 model would be 
different from the our non-SUSY model. We show its details in the
appendix \ref{sec:dsl}. 

iii) ~It is possible that a string could pass $r_m$ and approach to $r_0$ when it is pushed 
from a point $r_i (>r_m)$ toward $r_m$ with definite energy and velocity \cite{arXiv:0908.0407}.
However, it can never touch $r_0$ since an infinite energy is necessary to arrive there.  

iv) Another time dependent (moving) string is considered 
in the next section (Sec.4), namely the rotating closed string as a glueball state.
Its stable state with a finite energy and angular momentum is found at $r_m$, then it does not 
move from $r_m$ to the larger nor smaller $r$.


\vspace{.3cm}
\section{Glueballs as Rotating closed string}

Flavored mesons are given by a open string with two end points are on the D7 brane. 
On the other hand, the glueballs with higher spin would be represented by rotating closed strings in the bulk.
Such a rotating string is formulated as follows.

In the following analysis we adopt
the coordinate $z=R^2/r$, then the metric (\ref{finite-c-sol}) is written as follows
\beq\label{background3}
ds_{10}^2=e^{\Phi/2}{R^2 \over z^2}
\left\{A^2(z)\left(-dt^2+(dx^i)^2\right)+
   dz^2+z^2 d\Omega_5^2 \right\} \ .
\eeq 
Further, the metric for the string which rotates around
the $x_3$ axis is given by cylindrical polar coordinates as, 
\beq\label{background-susy}
ds^2_{(5)}=e^{\Phi/2}{R^2 \over z^2}\left(
 A^2(z)\left(-dt^2+d{\rho}^2+{\rho}^2d{\tilde{\theta}}^2+dx_3^2\right)+dz^2\right) \ . 
\eeq 
Taking the string world sheet as ${(\tau,\sigma)=(t,z)}$ and 
the ansatz, $\rho=\rho(z)$ and $\tilde{\theta}=\omega t$, the induced metric is 
given as
\beq
 {\cal G}_{\tau\tau}=e^{\Phi/2}{R^2\over z^2}A^2(z)\left(-1+\omega^2\rho^2\right)\, , \quad
 {\cal G}_{\sigma\sigma}=e^{\Phi/2}{R^2\over z^2}\left(A^2(z){\rho'}^2+1\right)\, ,
\eeq
where prime denotes the derivative with respect to $z$.
Then we have 
\beq\label{string-ac0}
 S_{\rm string}=\int dt {\cal L}=
     -{1\over 2\pi\alpha'}\int dtdz 
  e^{\Phi/2}A^2(z){R^2\over z^2}
        \sqrt{\left(1-\omega^2\rho^2\right)\left({\rho'}^2+A^{-2}(z)\right)}\, .
\eeq
From this, the spin $J_s$ and the energy $E_s$ of this string are given as
\beq\label{spin}
   J_s={\partial{\cal L}\over \partial\omega}
   ={1\over 2\pi\alpha'}\int dz e^{\Phi/2}A^2(z){R^2\over z^2}
        \omega\rho^2\sqrt{{\rho'}^2+A^{-2}(z)\over 1-\omega^2\rho^2}\, ,
\eeq
\beq\label{energy}
   E_s=\omega{\partial{\cal L}\over \partial\omega}-{\cal L}
   ={1\over 2\pi\alpha'}\int dz e^{\Phi/2}A^2(z){R^2\over z^2}
        \sqrt{{\rho'}^2+A^{-2}(z)\over 1-\omega^2\rho^2}\, .
\eeq
These are estimated by giving appropriate solutions for the corresponding
strings.

\vspace{.3cm}
\noindent{\bf Equations of motion for folded strings}

\vspace{.3cm}
In order to solve the string equation, it is convenient to use the reparametrization
invariant formalism since the 
configuration of a solution is given by a continuous curve. So the solution can be
expressed by one parameter, $s$ or $\sigma$ as given in \cite{GTT}.
The Lagrangian is written in terms of $s$ as
\beq\label{effL}
{\cal L}=-{1\over 2\pi\alpha'}\int ds {\tilde L}
     =-{1\over 2\pi\alpha'}\int_{s_i}^{s_f} ds 
  e^{\Phi/2}A^2(z){R^2\over z^2}
        \sqrt{\left(1-\omega^2\rho^2\right)\left(\dot{\rho}^2+\dot{z}^2A^{-2}(z)\right)}\, .
\eeq
where dot denotes the derivative with respect to $s$. $s_i(=0)$ and $s_f(=2\pi)$ are defined
as
\beq\label{bc-1}
  z(s_i)=z(s_f)\, , \quad \dot{z}(s_i)=\dot{z}(s_f)=0\, .
\eeq
\beq\label{bc-2}
  \rho(s_i)=\rho(s_f)\, , \quad \dot{\rho}(s_i)=\dot{\rho}(s_f)=0\, .
\eeq
These equations mean that the end point of the string is smoothly connected
since we consider closed string solutions. 

The equations of motion to be solved are obtained by introducing the canonical
momentum as,
\beq\label{eqs motion1}
  p_{\rho}={\partial {\tilde L}\over \partial\dot{\rho}}\, , \quad
  p_{z}={\partial {\tilde L}\over \partial\dot{z}}\, ,
\eeq
we have the Hamiltonian
\beq
  H=2{\tilde{H}\over \Delta}\, , \quad 
    \Delta={F\over \sqrt{\dot{\rho}^2+\dot{z}^2A^{-2}(z)}}\, ,
\eeq
\beq
  F= e^{\Phi/2}A^2(z){R^2\over z^2}
        \sqrt{1-\omega^2\rho^2}\, ,
\eeq
\beq
  \tilde{H}={1\over 2}\left(p_{\rho}^2+p_{z}^2A^2(z)-F^2 \right)\, .
\eeq

Then the Hamilton equations are obtained from $\tilde{H}$ instead of $H$
for the simplicity,
\beq\label{seq-1}
  \dot{\rho}=p_{\rho}\, , \quad \dot{z}=p_{z}A^2(z)\, , \quad 
\eeq
\beq\label{seq-2}
  \dot{p}_{\rho}=-\omega^2{\rho}Q^2(z)\, , \quad 
  \dot{p}_{z}=-p_z^2A(z){\partial A(z)\over \partial z}+
{1\over 2}\left(1-\omega^2\rho^2\right){\partial Q^2(z)\over \partial z}\, , \quad 
\eeq
and
\beq\label{Q-z}
  Q^2(z)=e^{\Phi}A^4(z){R^4\over z^4}\, .
\eeq

\vspace{.3cm}
\subsection {Solution in the SUSY background}

We solve above equations for the closed string in the
supersymmetric background,
(\ref{susy-solution}), by imposing the ansatz,
\beq\label{ansatz-z}
 z=z_m\, ,
\eeq
where $z_m$ is a constant. This satisfies the above boundary condition 
(\ref{bc-1}) of course. Then, from the second Eq.(\ref{seq-2}), we find
\beq
 z_m=\infty\, ,
\eeq
which means $r_m=R^2/z_m=0$. 

Alternative way to obtain this solution is as follows. First, rewrite the Eq.(\ref{effL}) 
by using ansatz (\ref{ansatz-z}) as follows 
\beq\label{effL-z}
{\cal L}=-{1\over 2\pi\alpha'}Q(z_m)\int d{\rho} 
        \sqrt{\left(1-\omega^2\rho^2\right)}\, ,
\eeq
where $Q(z)$ is given in the above (\ref{Q-z}). Then solving this with respect to $z_m$,
we find it as the minimum point of $Q(z)$. This point is already discussed above in the previous
section.

As for $\rho$, from the remaining equations
we find
\beq\label{c-sol2}
  \rho={1\over \omega}\sin\left({\sqrt{q}\over R^2}\omega s\right)\, .
\eeq
We use this simple slution in the followings.

\vspace{.5cm}
\noindent{\bf Regge behavior}

\vspace{.3cm}
The spin and the energy of this closed string configuration are given by using the
above equations (\ref{spin}) and (\ref{energy}) as
\beq\label{spin-2}
   J_s={1\over 2\pi\alpha'}2\int_{-1/\omega}^{1/\omega} d\rho 
      e^{\Phi/2}{R^2\over z^2}
        \omega\rho^2\sqrt{1+(\partial z/\partial \rho)^2\over 1-\omega^2\rho^2}\, ,
\eeq
\beq\label{energy-2}
   E_s=
   {1\over 2\pi\alpha'}2\int_{-1/\omega}^{1/\omega} d\rho  
      e^{\Phi/2}{R^2\over z^2}
        \sqrt{1+(\partial z/\partial \rho)^2\over 1-\omega^2\rho^2}\, .
\eeq
Substituting the above closed string solution, we find
\beq
  J_s={1\over 2\alpha'\omega^2}{\sqrt{q}\over R^2}\, , \quad
  E_s={1\over \alpha'\omega}{\sqrt{q}\over R^2}\, .
\eeq
Then we obtain
\beq
  J_s=\alpha'_{\rm glueball}E_s^2\, , \quad \alpha'_{\rm glueball}
   ={1\over 2}\alpha'{R^2\over \sqrt{q}}={1\over 2}{\alpha'\over Q(z_m)}
\eeq
Here we notice that
\beq
   \alpha'_{\rm glueball}={1\over 2}\alpha'_{\rm meson}
\eeq
where $\alpha'_{\rm meson}$ represents the slope parameter of the flavored 
mesons \cite{GTT}.

\vspace{.5cm}

\vspace{.5cm}
\noindent{\bf Problems of SUSY solutions}

We notice here that the above solution is pulled down upto $r=0$
by the gravitational attractive force. However, we find that $g_{rr}$ becomes infinite at this point.
Then it leads to a difficulty when we consider
quantum fluctuations around this classical configuration. 
We can expand the action around the above classical 
solution as
\beq
{\cal L}\equiv e^{\Phi/2}\sqrt{-g}=
\sqrt{q\frac{g_{0}}{r_{m}^{4}}}\left(1+\frac{1}{2g_{0}}\left\{
-\left(\rho^{\prime}\right)^{2}\delta\dot{r}^{2}+\left(1-\omega^{2}\rho^{2}\right)\delta
r^{\prime2}\right\} +\cdots \right)\, ,
\eeq
\beq
  g_0=(1-\omega^2\rho^2){\rho'}^2{r^4\over R^4}\Bigg{|}_{r_m}\, ,
\eeq
where dot and prime denote the derivative with respect to $\tau$ and $s$. The ellipsis
represent other fluctuations and higher order terms.
Then the coefficient of the quadratic terms of
$\delta r$ in ${\cal L}$ diverges like $1/r_m^4$ for $r_m=0$.
This implies that $\delta r$ must be suppressed, then the 
quantum fluctuation of the closed string configuration
given here cannot spread in the radial direction. 

This point is the defect of the present supersymmetric model. Consider the zero size limit of this closed string solution, then it corresponds to
a point particle in the bulk \cite{PRT}. It is dual to the glueball operator of 4D Yang-Mills
theory. However, this fluctuation could not propagate in the bulk.
As shown below, in the present case, we actually cannot find glueball spectra
through the fluctuations of the bulk fields in the supersymmetric bulk background.
This indicates that we must improve the background configuration such that 
the classical configuration of a closed string allows the quantum fluctuation
in the radial direction. One realization is given in the non-supersymmetric case
as shown below.

\vspace{.3cm}
\subsection {Solution for non-SUSY background}

For the non-supersymmetric background solution (\ref{non-susy-sol}), 
by solving the equation of motions (\ref{seq-1}) and (\ref{seq-2}), we find
$z_m$ as follows
\beq\label{c-sol1}
 R^2/z_m=r_m\approx 1.18~r_0\, .
\eeq
This is the same result with the one given in (\ref{rm}) since the same equation is solved. Namely
it is obtained as the minimum of $Q=\sqrt{|G_{tt}(r)|G_{ii}(r)}$. Then the closed strings are
trapped at $z_m$ and separated out of the singular point $r_0$.
This fact is different from the case of Witten's
D4-brane background because the closed strings are not trapped
at the horizon $U=U_{KK}$ in the background, and the strings 
continue to drop up to $U=0$ through the horizon $U_{KK}$.

As for $\rho(s)$, we find
\beq\label{c-sol3}
\rho=\frac{1}{\omega}\sin\left(Q(z_{m})\omega s\right)\, .
\eeq
While, in the supersymmetric case, the metric divergence has appeared at $r_m$, 
there is no such a metric divergence at $r_m$ in the present case, 
since $r_m>r_0>0$. 

We should notice here that $\rho$ is finite in spite of the fact that the string stays at
$r_m$. In the case of the open strings discussed above, the length becomes infinite
when the string approaches to $r_m$. There is no contradiction between the two results since
the closed string in the present case is rotating. In general, moving string could pass the point
$r_m$ if it has enough energy to climb the wall as seen in \cite{arXiv:0908.0407}

\vspace{.3cm}
\noindent{\bf Regge behavior}

The spin and the energy of this closed string configuration are estimated by using the
equations (\ref{spin}) and (\ref{energy}) as in the supersymmetric case. Then
we have the result,
\beq
  J_s={\alpha'}_{\rm glueball}^{\rm NS}E_s^2\, , \quad {\alpha'}_{\rm glueball}^{\rm NS}
   ={1\over 2}\alpha'{1\over {Q_m}}\, .
\eeq
Here we notice that $Q_m/(2\pi\alpha')$ represent the tension of the 
quark and anti-quark linear potential obtained for the non-supersymmetric model
used here, then we also find
\beq
   {\alpha'}_{\rm glueball}^{\rm NS}={1\over 2}{\alpha'}_{\rm meson}^{\rm NS}
\eeq
for mesons with large spin. 

\section{Glueballs from bulk field fluctuations}

\subsection{non-Supersymmetric case}
For non-supersymmetric background, we find the classical
stable configuration of glueballs corresponding to the large quantum number
state. And quantum fluctuations can be added them to see the corrections to the
Regge behavior obtained above. For zero size limit of the classical string, namely
the point particle case,
we study the corresponding
glueball state by solving the field equation of the quantum fluctuation
of the bulk fields as given below.

\vspace{.5cm}
\noindent{\bf  Graviton $2^{++}$; }
As for the glueball spectrum, many attempts have been made by solving the
linearized field equations of bulk field fluctuations in the given background.
Here we consider the field equation of the traceless and transverse component
of the metric fluctuation, which is denoted by $h_{ij}$. 
{
Its linearized equation is given 
in the Einstein frame metric as 
\beq\label{graviton-eq}
  {1\over\sqrt{-g}}\partial_M\left(\sqrt{-g}g^{MN}\partial_N h_{ij}\right)=0\,,
\eeq
where we use $z$ instead of $r$ and assumed as $h_{ij}=h_{ij}(x^0,x^i,z)$,} then
$M,N$ are the five dimensional ($(x^0,x^i,z)$) suffices.\footnote{In the string frame metric case, this equation is written as
${1\over\sqrt{-g}}\partial_M\left(\sqrt{-g}
e^{-2\Phi}g^{MN}\partial_N h_{ij}\right)=0$ as given in \cite{Minahan}}
This equation
is equivalent to the massless scalar field equation. As shown in \cite{BMT},
this equation is common to $2^{++},~1^{++}$ and the one of the non-active 
\footnote{Here active means that the dilaton background solution is nontrivial
as in the present case.}
dilaton $0^{++}$,
which are dual to the glueball of $F_{\mu\nu}F^{\mu\nu}$.
While it is usually used to derive the type IIA theory, the NS-NS part
is common with the one of the type IIB theory. 
Then the masses of these three spin states degenerate. However, we are considering
non-trivial dilaton background configuration, then the above Eq.(\ref{graviton-eq})
is used for the graviton fluctuation, the glueball of $2^{++}$ state.

By setting as $h_{ij}=p_{ij}e^{ikx}\phi(z)$
and $-k^2=m^2$, we get\footnote{
$p_{ij}$ denotes projection operator onto the traceless and transverse components.
}
{
\bea
   &&\partial_z^2\phi+g_z(z)\partial_z\phi+{m^2\over A^2 }\phi=0\, ,\label{glueball-eq-1}\\
  && g_z(z)=\partial_z\left(\log\left[(R/z)^3A^4\right]\right)=-{3\over z}+4{\partial_zA\over A}\, .
\label{glueball-eq-2}
\eea}
Here we notice
\beq
 A=\left(1-\left({z\over z_0}\right)^8\right)^{1/4}
\eeq
where $z_0=R^2/r_0$. Then we see that
the equation (\ref{glueball-eq-1}) has 10 regular singularities at
$z=0,~\infty$ and the points of $\left({z\over z_0}\right)^8=1$.
We therefore try to find the eigenfunctions in the region of $0\leq z\leq z_0$ through
WKB approximation \cite{Minahan,COOT}
by changing the variable from $z$ to $y$ which is defined as
\beq
 z={z_0\over 1+e^{y}}\, , \quad z_0={R^2\over r_0}
\eeq
where $y$ is defined in the region of $-\infty<y<\infty$. 

\vspace{.3cm}
Then the equation (\ref{glueball-eq-1})
and (\ref{glueball-eq-2}) are rewritten as
\bea\label{glueball-eq-y}
   &&\partial_y^2\phi+g_2(y)\partial_y\phi+{m^2\over A^2 }
         {z_0^2e^{2y}\over (1+e^y)^4}\phi=0\, ,\label{glueball-eq1-y}\\
  && g_2(y)={5e^y\over 1+e^y}-1+4{\partial_yA\over A}\, . \label{glueball-eq2-y}
\eea

\begin{figure}[htbp]
\vspace{.3cm}
\begin{center}
\includegraphics[width=12cm]{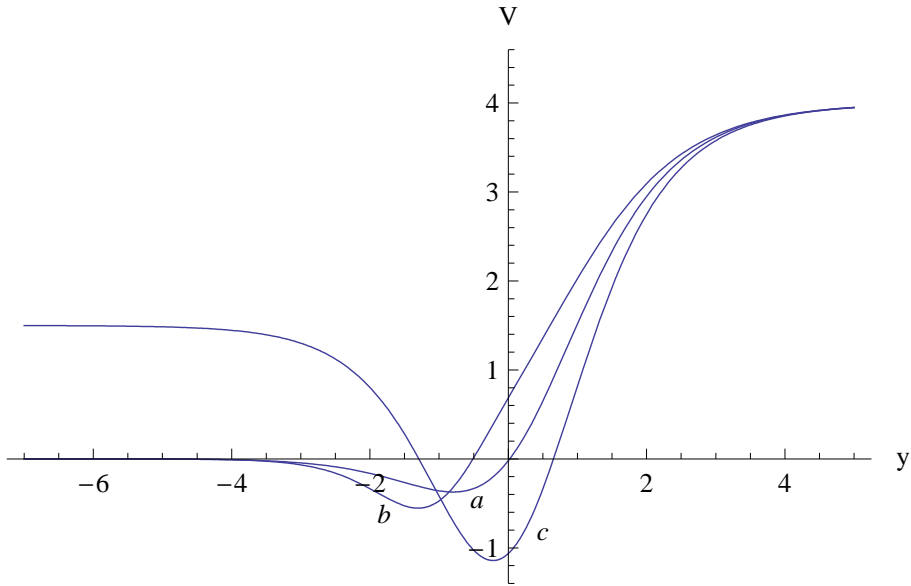}
\caption{{\small The Schr\"{o}dinger potentials $V(z)$ with $z_0=2$ (GeV$^{-1}$)
are shown
for (a) graviton $m=2.182 (2.15)$(GeV) , (c) axion $m=3.05 (2.25)$(GeV) ,  and (b) dilaton $m=1.207 (1.47)$(GeV)  cases. The values in the parenthesis are the data of the lattice simulation \cite{Colin,Chen,Me}.
}}
\label{NS-Pot}
\end{center}
\end{figure}

\vspace{.2cm}
In order to perform the WKB approximation, we rewrite the wave function
as $\phi=e^{-{1\over 2}\int dy g_2(y)}f(y)$, then we obtain
\beq\label{glueball-eq2}
  -\partial_y^2f+V(y)f=0\, , \quad {V={1\over 4}g_2^2+{1\over 2}\partial_yg_2-
{m^2\over A^2 }{z_0^2e^{2y}\over (1+e^y)^4}\, }.
\eeq
This is the one dimensional Schr\"{o}dinger equation form with the potential $V$
and the zero energy eigenvalue. For an appropriate mass $m$,
we can see that $V$ has two turning points, $y_1$ and $y_2(>y_1)$, to give \cite{Minahan}
\beq\label{WKB-condi}
  \int_{y_1}^{y_2}\sqrt{-V}dy=\left(n+{1\over 2}\right)\pi
\eeq
with integer $n$. From this equation we obtain the discrete glueball mass $m_n$, where
$n$ denotes the node number of the eigenfunction. The potential for the zero node is
shown in the Fig.\ref{NS-Pot}.

\vspace{.3cm}
Here we should give the following comments.

{\textbullet} We notice here that the two turning points found above are finite. This fact
is understood as follws. The "Schr\"{o}dinger" potential $V(y)$ given in (\ref{glueball-eq2})
is expanded for small $e^y(\equiv x)$ as
\beq
   V(y)={x\over 4} - { x^{3/2}\over 2 \sqrt{2}}(mz_0)^2 + {77 x^2\over 16}+O(x^{5/2})\, .
\eeq
This implies that $V(y)$ change sign near $x=0$ (at about $x\sim 1/(2m^4z_0^4)$). 
On the other hand, at large $x$, we have
\beq
  V(y)=4 - {15\over 2 x} + \left({45\over 4} - (mz_0)^2\right) {1\over x^2}+O(1/x^3)\, .
\eeq
Then, $V(y)$ approaches to $4(>0)$ at large $x$. Therefore, there are two turning points
at finite $y$ or $x$. This point is very important since the turning point in the smaller
side of $y$ is found at $y=-\infty$ in the D4 model \cite{Minahan} and Constable-Myers model
\cite{CoM}. In the Fig. \ref{NS-Pot}, the potential $V(y)$ is shown for the zero node state, and
we actually have the two finite turning points in our case as, $y_+=0.02065$ and $y_-=-6.526$.

{\textbullet} In the case of D4 model, the point $y=-\infty$ corresponds to the event horizon of the
bulk black hole background. This point, fortunately, is not a singular point of the supergravity
background, then it would be meaningful to impose a boundary condition as a turning
point for the WKB approximation. However, for the Constable-Myers model, this point $y=-\infty$
is at the naked singularity of the bulk background, so the authors of \cite{CoM} concluded as
that the glueballs of $2^{++}$ and $0^{++}$ cannot be seen in their model due to this
reason. 

{\textbullet} In our model, the turning points are far from $y=-\infty$, where the naked
singularity exists, then we can perform the WKB analysis without worrying on this point. This would
be reduced to the fact the singular point is at $r=r_0$, which is finite, and the wall, which
push out the classical string configurations and D-branes, also at the outside of the
singularity.
We suppose that, due to this wall, the quantum wave-function is also confined
in a finite range, then we could find discrete spectrum as shown below in terms of the
WKB approximation.

\vspace{.5cm}
\noindent{\bf Axion $0^{-+}$; }
Since the axion and three form field strengths, which couple to the axion,
are non-active in the present background, then the equation of motion
for the axion fluctuation is obtained directly from the bulk action as
\beq\label{axion-eq}
  {1\over\sqrt{-g}}\partial_M\left(\sqrt{-g}g^{MN}e^{2\Phi}\partial_N \chi\right)=0\,,
\eeq
As above, we rewrite this equation using the following form, 
$\chi=e^{ikx}e^{-{1\over 2}\int dy g_{\chi}(y)}f_{\chi}(y)$, as follows
\beq\label{glueball-eq3}
  -\partial_y^2f_{\chi}+V_{\chi}(r)f_{\chi}=0\, , \quad 
  V_{\chi}={1\over 4}g_{\chi}^2+{1\over 2}g_{\chi}'-{m^2\over A^2}{z_0^2e^{2y}\over (1+e^y)^4}\, .
\eeq
where
\beq
 g_{\chi}(y)=-g_2(y)+2\partial_y\Phi \,
\eeq
For
$m =3.05$GeV and $z_0 = 2.0$GeV$^{-1}$, the values of the potential $V_{\chi}$ is shown in the Fig.\ref{NS-Pot}.
As shown in this figure, for the case of the axion,
we could find two turning point (zero-point) for large enough
value of $m$, then the WKB method is useful as in the 
graviton case.

\vspace{.5cm}
\noindent{\bf Dilaton $0^{++}$; }

In the present case, the dilaton is an active scalar, namely it has a classical
configuration. Then its fluctuation $\phi$ mixes with the scalar component of the graviton.
Expanding the metric in terms of the scalar $\psi$ and traceless transverse part $h_{ij}^{TT}$
as
\beq
 h_{ij}=2\eta_{ij}\psi+h_{ij}^{TT}+\cdots\, .
\eeq
Then the equation of motion of the scalar mode $\zeta$, which is
invariant under 
the general coordinate transformation, is given as \cite{Kiritsis:2006ua}
\beq
  \partial_y^2\zeta+g_{\zeta}(y)\partial_y\zeta+{m^2\over A^2}{z_0^2e^{2y}\over (1+e^y)^4}\zeta=0\, ,
\eeq
\beq
   g_{\zeta}(y)=g_2(y)+2{\partial_yB \over B}\, ,
\eeq
\beq
 \zeta=\psi-\phi B\, ,
 \quad B={\partial_y\Phi\over \partial_y\left(\log({R(1+e^y)\over
 z_0}A)\right)}\, .
\eeq 

As above, we rewrite this equation by using 
$\zeta=e^{ikx}e^{-{1\over 2}\int dy g_{\zeta}(y)}f_{\zeta}(y)$, as follows
\beq\label{glueball-eq4}
  -\partial_y^2f_{\zeta}+V_{\zeta}(r)f_{\zeta}=0\, , \quad 
  V_{\zeta}={1\over 4}g_{\zeta}^2+{1\over 2}\partial_yg_{\zeta}-{m^2\over A^2}{z_0^2e^{2y}\over (1+e^y)^4}\, .
\eeq
Then we can perform the WKB analysis as above.
{The behavior of the potential $V_{\zeta}$ is similar to the one of the graviton
as seen from Fig. \ref{NS-Pot}. However, due to a slight difference of the potentials leads to
the difference of the eigen masses for the graviton and the dilaton as shown in the 
Table \ref{simulation}.
In any case, the infinite series of the radial exitations for the three states are observed
in the case of the non-supersymmetric background solution.}

\subsection{Supersymmetric case}

In the supersymmetric case, we cannot find glueball state
from the fluctuation mode of the bulk fields since there is no normalizable wave
function with definite four dimensional mass eigenvalue. 

\vspace{.5cm}
\noindent{\bf Graviton $2^{++}$; }

Firstly, this is shown for the
graviton fluctuation. Its equation is given by (\ref{graviton-eq}), but the metric is
used for the supersymmetric solution. In this case, there is no restriction to the variable $r$
and we can consider the whole range, $0\leq z<\infty$. The equation for the graviton is given 
by setting as above, $h_{ij}=p_{ij}e^{ikx}\phi_s(z)$
and $-k^2=m^2$, then we get
\beq\label{graviton-susy}
   \partial_z^2\phi_s-{3\over z}\partial_z\phi_s+{m^2}\phi_s=0\, .
\eeq
This is solved as
\beq
  \phi_s(z)={z^2\over z_0^2}\left( C_1J_2(mz)+C_2N_2(mz)\right)
\eeq
where $C_{1,2}$ are arbitrary constant and $m$ denotes the glueball mass. $J_n(x)$ and
$N_n(x)$ are the first kind and second kind Bessel functions. It is easily assured that
this solution is not normalizable since
\beq
  \int_0^{\infty}{dz\over z^3}|\phi_{s}(z)|^2
\eeq
is divergent. This is because of that there is no infrared cutoff or wall in this case.

\vspace{.3cm}
From the viewpoint of the one-dimensional Schr\"{o}dinger equation, we can see 
that the potential has no two turning points. Actually, for the graviton 
we have
\beq
  V_{gr}={15\over 4z^2}-{m^2}\, .
\eeq
{This potential has only one zero point for any $m$, then we could not obtain
any mass state by the WKB approximation by using this potential.}

\vspace{.5cm}
\noindent{{\bf Dilaton and Axion $0^{\pm +}$; }}

\vspace{.3cm}
{In the present case, both the dilaton and the axion have its classical
solution. Then their fluctuations mix with the gravitational ones. Here we consider the
mass eigenmodes of the two scalar fluctuations in a special gauge, where they decouple
from each others. This is performed as follows.}

{At first, set the flucuations, $h_{MN}$, $\delta\Phi$, and $\delta\chi$, of each field as,
\beq
 g_{MN}=a^2(z)(\eta_{MN}+h_{MN})\, , \quad \Phi=\bar{\Phi}+\delta\Phi\, , \quad \chi=\bar{\chi}+\delta\chi
\eeq
where $\bar{\Phi}$ and $\bar{\chi}$ denote classical solutions,
\beq
\bar{\chi}=-e^{\bar{\Phi}}+\chi_0\, , \quad e^{\bar{\Phi}}=1+\tilde{q}z^4\, , \quad
\tilde{q}=q/R^8
\eeq
Then, from (\ref{p5d}) and (\ref{chieq5d}), we obtain
\bea\label{p1}
 && {1\over a^3}\partial_M\left(a^3\partial^M\delta\Phi+a^3
    \left({1\over 2}h\eta^{MN}-h^{MN}\right)\partial_N\bar{\Phi}\right)= \nonumber \\
 && - \left({1\over 2}h\eta^{MN}-h^{MN}\right)\partial_M\bar{\Phi}\partial_N\bar{\Phi}
 -2\left(e^{\bar{\Phi}}\partial_M\delta\chi+\delta\Phi\partial_M\bar{\Phi}\right)\partial^M\bar{\Phi}
\eea
\beq\label{chieq2}
 \partial_M\left(a^3e^{2\bar{\Phi}}\eta^{MN}\partial_N\delta\chi+a^3
\left({1\over 2}h\eta^{MN}-h^{MN}+2\delta\Phi\eta^{MN}\right)e^{\bar{\Phi}}\partial_N\bar{\Phi}\right)=0
\eeq
where $$ h=h^M_M$$ and
the suffices $M,N$ are raised and lowered by $\eta_{MN}$ or $\eta^{MN}$.}

{Here we take the following gauge conditions,
\beq\label{gc-1}
   {1\over 2}h-h^{zz}+2\delta\Phi=0
\eeq
\beq\label{gc-2}
    \partial_{\mu}h^{\mu z}-2e^{\bar{\Phi}}\partial_z\delta\chi=0
\eeq
and 
\beq\label{vector-gauge}
 f_{\mu}=0\, , 
\eeq
where $f_{\mu}=0$ is defined as
\beq\label{met-fl-3}
h_{MN}= \left(\matrix{2 h_{\mu\nu} 
+(\partial_{\mu} f_{\nu} +\partial_{\nu} f_{\mu}) 
+ 2\eta_{\mu \nu} \psi
+ 2 \partial_{\mu}\partial_{\nu} E
& B_{\mu} + \partial_{\mu} C &\cr
B_{\mu} + \partial_{\mu} C  & 2 \xi &\cr}\right),
\eeq
where we used the same setting with \cite{Gio}.
Then, (\ref{gc-1}) and (\ref{gc-2}) are rewritten 
in terms of the fields in (\ref{met-fl-3}) as
\beq\label{gc-11}
   2\Psi-\xi+\partial_{\alpha}^2E+2\delta\Phi=0\, ,
\eeq
\beq\label{gc-22}
    \partial_{\mu}^2 C+2e^{\bar{\Phi}}\partial_z\delta\chi=0\, .
\eeq
As a result, the gauge is completely fixed by the above three conditions.
}

\vspace{.3cm}
{Then the eqautions (\ref{p1}) and (\ref{chieq2}) are rewritten as
\beq\label{p2}
  \delta\Phi''+3{a'\over a}\delta\Phi'+\left(m^2_{\Phi}-4(\bar{\Phi}')^2\right)\delta\Phi=0
\eeq
\beq\label{chieq3}
  \delta\chi''+\left(3{a'\over a}+4\bar{\Phi}'\right)\delta\chi'+m^2_{\chi}\delta\chi=0
\eeq
where prime denotes the derivative with respect to $z$ and
$$ \partial_{\mu}^2\delta\Phi=m_{\Phi}^2\delta\Phi$$
$$ \partial_{\mu}^2\delta\chi=m_{\Phi}^2\delta\chi$$}

{Then the equation (\ref{p2}) is rewritten by using
$$ \delta\Phi=e^{ikx}e^{-{1\over 2}\int dz g_{\phi}}f_{\phi}(z)  $$
as
\bea
  &&-f_{\phi}''+V_{\phi}(z)f_{\phi}=0\, , \\
  && V_{\phi}(z)={1\over 4}g_{\phi}^2+{1\over 2}g_{\phi}'-m_{\phi}^2-4(\bar{\Phi}')^2\, \\
  && g_{\phi}=3{a'\over a}=-3{1\over z}
\eea
And by using
$$ \delta\chi=e^{ikx}e^{-{1\over 2}\int dz g_{\chi}}f_{\chi}(z)  $$
(\ref{chieq3}) is rewritten as
\bea
  &&-f_{\chi}''+V_{\chi}(z)f_{\chi}=0\, , \\
  && V_{\chi}(z)={1\over 4}g_{\chi}^2+{1\over 2}g_{\chi}'-m_{\chi}^2\, \\
  && g_{\chi}=3{a'\over a}+4\bar{\Phi}'
\eea}

\begin{figure}[htbp]
\begin{center}
\includegraphics[width=7.0cm,height=6cm]{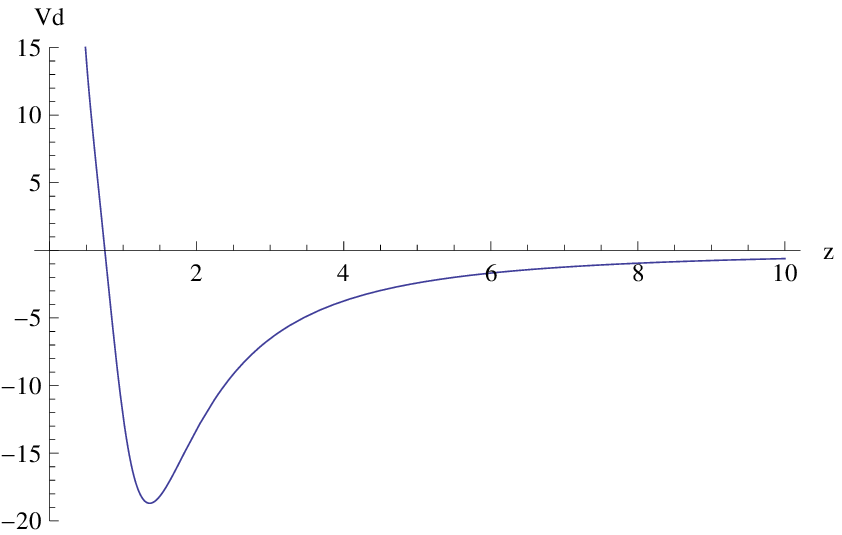}
\includegraphics[width=7.0cm,height=6cm]{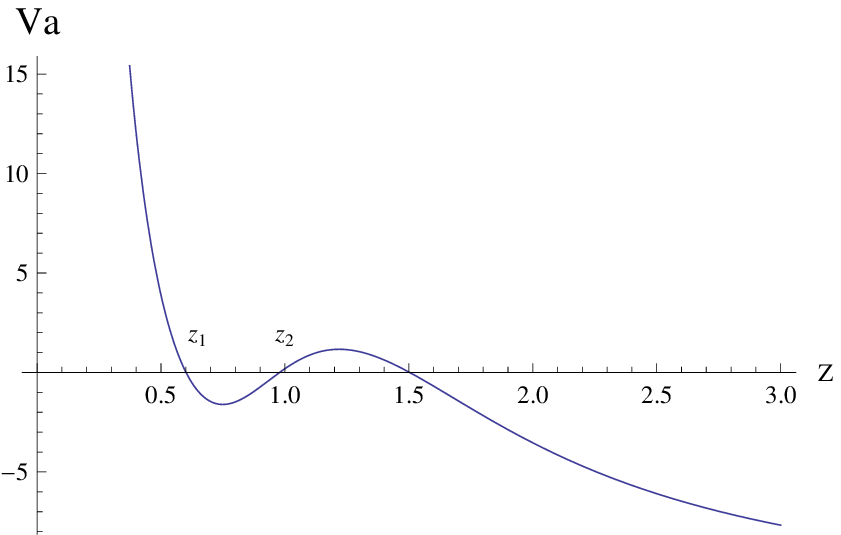}
\caption{Shrodinger potential for the dilaton (left) and axion (right) in SUSY solution $q=1$, $m_{\phi}=0.10$,
and $m_{\chi}=3.4$.
\label{P-dilaton}}
\end{center}
\end{figure}

{The typical potentials are shown for both cases
in the Fig. \ref{P-dilaton}. As for the dilaton, 
the potential has a deep negative minimum, but the second zero point does not appear at the 
large z side. This situation
is therefore similar to the one of the graviton, in which case there is no minimum.
Namely only one zero point is observed
for any $m_{\phi}$. Then we can not find any glueball state with a finite mass.}

{As for the case of axion,
on the other hand, there seems to be a possibility of glueball's existence
due to the minimum of the potential and two zero points, say $(z_1,z_2)$ (see Fig. \ref{P-dilaton}),
which are seen in the Fig. \ref{P-dilaton}. However, this minimum is not deep enough
to produce a glueball state. We examined the value of
$$ \int_{z_1}^{z_2}\sqrt{-V_{\chi}(z)}dz$$ for varous parameter ranges of $q$ and $m_{\chi}$,
but it is too small and does not satisfy the condition needed for the WKB bound state or (\ref{WKB-condi}).
}

{As a result, the two scalar modes also have no glueball state as the graviton.}
This fact can be related to
the metric singularity as mentioned above. Due to this singularity, the fluctuation
of the classical closed string configuration cannot spread in the bulk.
There might be several directions to remove this difficulty.
One easy way is to introduce an artificial cutoff for the coordinate
$r$. Another would be a modification of the model to our non-supersymmetric case
as a simple example, which is shown above.

\section{Numerical results for the glueball mass} 

\begin{figure}[htbp]
\vspace{.2cm}
\begin{center}
\includegraphics[width=7cm]{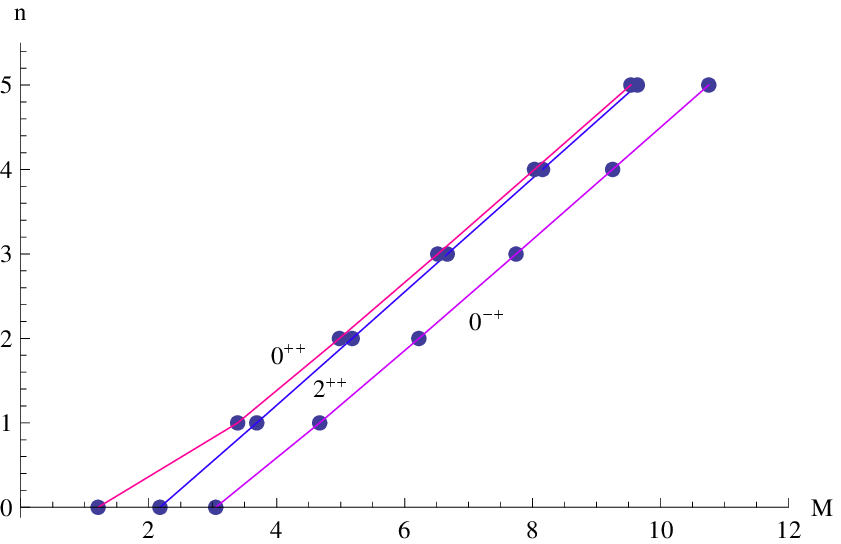}
\includegraphics[width=7cm]{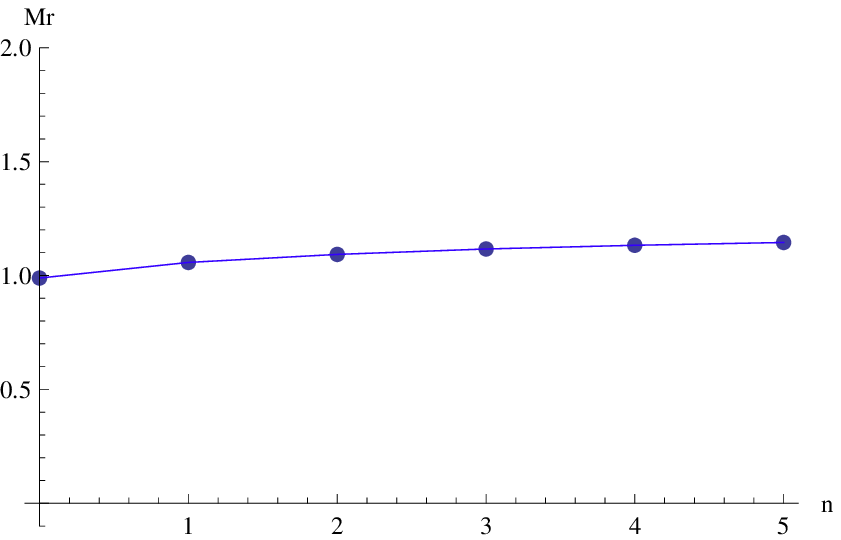}
\caption{{\small {\bf Left;} Numerical results of our the glueball mass for ${r_0\over R^2}=0.5$. 
$n$ denotes the node number
of the states. {\bf Right;} The mass ratio, (our calculation)/(the one of D4 model \cite{BMT}),
for $J^{PC}=2^{++}$ spectra.}}
\label{WKB}
\end{center}
\end{figure}

The glueball mass depends only on the parameter $r_0/R^2$ in our model. We show our results 
in the Fig. \ref{WKB} and in the Table \ref{simulation}. 

\begin{table}[htbp]
\begin{center}
\caption{{\small The glueball masses for $r_0/R^2=0.5 ~(\mbox{GeV})$. 
The column WKB shows our result of WKB calucuations 
in the unit of GeV. ${J^{PC}}_n$ denotes spin ($J$), 
charge conjugatin ($C$), parity ($P$),
and node number ($n$) of the corresponding wave-functions respectively.}}
\vspace{.2cm}

\begin{tabular}{|l|c||l|c||l|c|}
\hline 
${J^{PC}}_n$ &  WKB & ${J^{PC}}_n$  & WKB   & ${J^{PC}}_n$ & WKB \\ \hline
${2^{++}}_0$ & 2.176 &${0^{-+}}_0$  & 3.049 &${0^{++}}_0$ &  1.207 \\ \hline
${2^{++}}_1$ & 3.689 &${0^{-+}}_1$  & 4.673 & ${0^{++}}_1$ &   3.390 \\ \hline
${2^{++}}_2$ & 5.181 &${0^{-+}}_2$  & 6.221 &${0^{++}}_2$ & 4.981 \\ \hline
${2^{++}}_3$ & 6.668 &${0^{-+}}_3$  & 7.743 &${0^{++}}_3$ & 6.516 \\ \hline
${2^{++}}_4$ & 8.154 &${0^{-+}}_4$  & 9.251 &${0^{++}}_4$ & 8.031 \\ \hline
${2^{++}}_5$ & 9.639 &${0^{-+}}_5$  & 10.752&${0^{++}}_5$ & 9.535 \\
\hline
\end{tabular}
\label{simulation}
\end{center}
\end{table}

They are obtained by using $M_0(2^{++})=2.176$, which is given here as an average of
the lattice simulation \cite{Me,Colin,Chen} and used as an input data. 
To use this value as an input is equivalent to fix the parameter of our model as
\beq\label{para-1}
  {r_0\over{R^2}}=0.50~ \mbox{GeV}\, .
\eeq
Our results approximately reproduce the other data of the lattice simulation
given as \cite{Me} $M_0(0^{++})=1.475$ and $M_0(0^{-+})=2.25$ (GeV) for the lowest modes.
The masses of the exited state with higher node are also
shown. Those one of $2^{++}$ are compared with the results obtained in a different holographic
model \cite{BMT}, and we could find that they are almost equal each other. For other spin states,
which are not shown here, the ratios for those spectra are similaly near one.

\vspace{.3cm}
On the other hand, we know another simulation result for the gauge condensate, 
$\langle F_{\mu\nu}F^{\mu\nu}\rangle $, which is given as \cite{EGM}
\beq\label{vev}
  {\lambda\over 4\pi^2}\langle F_{\mu\nu}F^{\mu\nu}\rangle=0.14 \mbox{GeV}^{4}\, .
\eeq
When this is used, the parameter $r_0/R^2$ can be determined independently of the data
for any glueball mass. In our model, there are three independent parameters, 
$r_0$, $R$, and $\lambda$. They are fixed as follows;

\vspace{.3cm}
The value of $r_m$, which is defined as the minimum of the 
gravitational potential $Q(r)$, is given as
\beq
  {R^2\over z_m}=r_m=1.18r_0
\eeq
with
\beq\label{rm2}
  Q(r_m)\equiv Q_m=2.40\left({r_0\over R}\right)^2\, . 
\eeq
Then the meson Regge slope parameter is written in terms of this $Q_m$ as
\beq
   \alpha'_{Meson}={\alpha'\over Q_m}
  ={\sqrt{\lambda}\over 2.40}\left({\alpha'\over r_0}\right)^2\, .
\eeq
Next, expanding the dilaton as
\beq
  e^{\Phi}=1+\sqrt{6}{r_0^4\over r^4}+\cdots\, ,
\eeq
we find
\beq
 q=\sqrt{6}r_0^4=\pi^2\langle
 F_{\mu\nu}F^{\mu\nu}\rangle\lambda{\alpha'}^4\, .
\eeq
Then, using (\ref{vev})
we obtain
\beq
  {r_0\over{\alpha'}}=2.17 \mbox{GeV}\, .
\eeq
And, from (\ref{rm2}) we find
\beq
    \alpha'_{Meson}=0.088\sqrt{\lambda} ~(\mbox{GeV}^{-2})\, .
\eeq
Then, finally we get
\beq\label{para-2}
  {r_0\over R^2}={r_0\over \alpha'\sqrt{\lambda}}={0.191\over \alpha'_{Meson}}
    \sim 0.218 ~{\rm GeV}\, .
\eeq
When we respect this result, we find that the glueball masses are half of the one
obtained by using (\ref{para-1})


\vspace{.4cm}
Both results (\ref{para-1}) and (\ref{para-2}) are obtained by 
using the lattice simulations as the input in our
analysis, and they are not compatible. 
If both the two lattice results are correct, this implies that our model is so simple that
we could not reproduce well the lattice results since the number of the parameters of our theory
may be too small. We should add other freedom in our theory, for example other 
bulk field condensations should be considered, but it is not our present
scope to discuss this point.

\section{Summary and Discussion}

We have studied the role of the dilaton field being played
in the holographic hadron physics, especially
for the glueballs. Here we have studied
two dilaton configurations, for supersymmetric and non-supersymmetric cases. In both cases,
the dilaton contains the condensate of the
gauge field strength, $\langle F_{\mu\nu}F^{\mu\nu}\rangle$, which determines 
the properties of the vacuum 
of the Yang-Mills theory. This condensate is intimately related to
the quark confinement and the tension of the linear potential between the quark and the anti-quark. 
This has been assured through the study of the 
Wilson-loop and classical string configurations obtained as solutions of the Nambu-Goto
action.

The analysis is extended here to the glueballs (closed string), and we find the Regge behavior
through the classical solutions 
for the folded closed string case. The result shows the slope of the glueball trajectory
is the half of one of the flavor mesons which are given by the open strings. This relation is
expected from the configurations of the folded closed string which has two times length of the 
extended part of the long open string. This behavior is seen both in the supersymmetric and 
non-symmetric cases.

In the supersymmetric case, however, the stable closed string (classical configuration) 
is found at the horizon of the background ($r=0$) where the metric singularity
is seen in the radial coordinate $r$ direction. This implies that
the fluctuation of the closed string in the $r$ direction should be suppressed.
Then we cannot expect the quantum fluctuation mode of the glueballs in the bulk which
extends to the direction $r$. Actually, this point is assured by solving directly the equations of
motion of the quantum fluctuations of the bulk fields. As expected, we cannot find any
glueball state in this case. 

In order to evade the metric singularity mentioned above, it would be needed to introduce
an appropriate infrared cutoff in the theory. 
Although it is easy to introduce it by hand as in the hard wall
model \cite{EKSS,TB}, instead, we move to the non-supersymmetric solution to find a
wall which supports the glueball states.

\vspace{.3cm}
For non-supersymmetric case, the bulk curvature is singular at $r_0$, but 
there is no metric singularity at 
the position, $r=r_m (>r_0)$, where the classical closed-string is obtained. The reason why the
classical solution is trapped at this point is that this point is the minimum point of
the gravitational potential for the strings. As for the open string, its energy increases
with increasing angular momentum or spin. As a result, it grows long and the prolonged
part approaches to $r=r_m$. However it cannot go over this point $r_m$ even if the
energy becomes infinite. In this sense, 
the static strings are blocked there due to the wall.

Furthermore, we need probe branes, D7 branes for flavored quarks and D5 branes for baryon vertex,
to construct hadrons. When the branes are embedded in the non-supersymmetric background, they are
also blocked by the same wall and cannot penetrate into the region $r<r_m$. Namely their embedded configurations are 
stabilized in the region of $r>r_m$.

\vspace{.3cm}
As for the small spin states, they are expressed by the quantum fluctuations on the probe brane
for the flavor mesons or in the bulk for the glueballs. In the case of the flavor mesons,
their mass gap is given by the quark mass
$m_q=w(\infty)$ for the supersymmetric case. However, in the non-supersymmetric case, the
wall generates an infrared cutoff $w(0)(>r_m)$ due to the finite chiral condensate even if
$m_q=0$. So we could find flavored mesons with finite mass 
in the non-supersymmetric case even if $m_q=0$.

In the case of the glueballs, on the other hand, the quarks are not contained in the state,
then there is no mass scale to give a mass gap for the supersymmetric case. Then we cannot find any normalizable wave-functions of the glueball wave equation. 
On the other hand, in the non-supersymmetric case, there appears a wall
near $r_0$, where infinite high potential wall stands for the strings. This implies that we should
find the solutions of the glueball wave-equations by restricting the dynamical region of the
wave function to $r_0<r$. This procedure is also adopted in the Witten model, in which the bulk
configuration has a horizon coming from the bulk black hole geometry. So this point is not a 
singular point of the curvature, but metric singularity is generated at this point. In order
to evade this singularity, the region of the radial variable is restricted above the horizon
by introducing an appropriate change of the variable. This procedure is equivalent to introduce
an infrared cutoff, which provides a mass gap for the glueball.

In this case, we find discrete spectra of glueballs in the non-supersymmetric case.
It is possible to adjust the parameters of the theory consistently with the 
lattice-simulation results for the glueball masses. On the other hand, we know the lattice data
for the gauge condensate, $\langle F_{\mu\nu}F^{\mu\nu}\rangle$. When we respect this data,
however, we find about half values of the glueball mass which are given in the lattice simulation.
How to reconcile these two lattice-results with our model is an open problem here.

\vspace{.5cm}
As for the confining theory proposed by Klebanov and Strassler\cite{KSt}, it is supersymmetric and the
potential $Q(\tau)$ \footnote{In the model of Klebanov and Strassler, the variable $\tau$
corresponds to our $r$.}
has a finite minimum at $\tau=0$, then there is no infrared cutoff as in our 
non-supersymmetric case.
However, in the case of KS model, there is no metric singularity at $\tau=0$ so the quantum
corrections around this minimum point are calculable for the classical closed-string solutions.
Reflecting this fact, the glueball spectra are obtained without introducing 
any infrared cutoff. The dual gauge theory of this model is however different from our's
since the gauge condensate is not present in this case. It would be an interesting problem
to study the relation to the dual theory of our model. This would be an open problem here.

\vspace{.3cm}
\section*{Acknowledgments}
{K. Ghoroku thanks to Hirofumi Kubo 
for useful discussions in several parts of the contents.}



\def\theequation{A. \arabic{equation}}
\setcounter{equation}{0}

\appendix

\section{The evaluation of the U-shaped string length} \label{sec:dsl}

\begin{figure}[htbp]
\vspace{.3cm}
\begin{center}
\includegraphics[width=7cm]{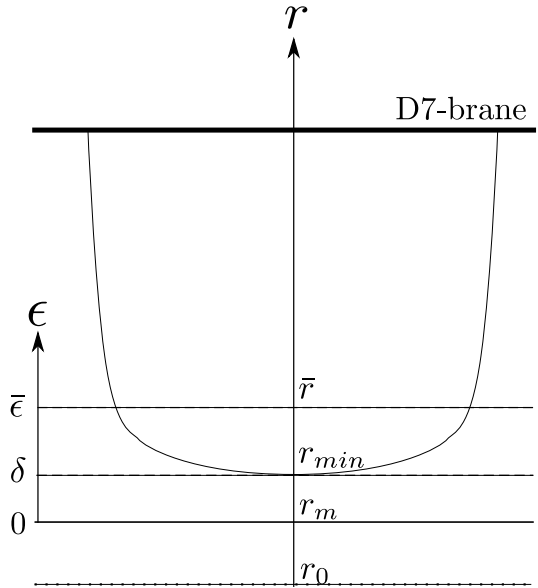}
\caption{{\small The relations between variables. $\epsilon$
 measures a interval from $r_m$ to $r$ and it can take a value from
 bottom point
 $\delta$ to some finite value $\bar{\epsilon}$}}
\label{variables}
\end{center}
\end{figure}

Here we show that the energy configuration of U-shaped open-string,
which corresponds to Wilson loop in the dual gauge theory, is infinite at
$r=r_m$. 
To show that, we evaluate the first integral in (\ref{L-sep-int}) by
changing integral variable as $\epsilon \equiv r-r_m$ and setting $\delta
\equiv r_{min}-r_m$. Relations between variables are
 illustrated in Fig. \ref{variables}.
Expanding the function in the root in denominator of this integrand
around $r=r_m$ up
to second order respect to $\epsilon$ or $\delta$, it
can be evaluated as follows,
\begin{eqnarray}
I(\bar{\epsilon},\delta)&\equiv&\int_{\delta}^{\bar{\epsilon}}d\epsilon\frac{Q(r_{min})}{\sqrt{H(r)\left(Q^{2}(r)-Q^{2}(r_{min})\right)}}\nonumber \\ 
& = & \int_{\delta}^{\bar{\epsilon}}d\epsilon\frac{Q(r_{min})}{\sqrt{a\epsilon^{2}+b\epsilon+c}}\nonumber \\
 & = & \int_{\delta}^{\bar{\epsilon}}d\epsilon\frac{Q(r_{min})}{\sqrt{a\left(\epsilon+\frac{b}{2a}\right)^{2}-\frac{b^{2}}{4a}+c}}\nonumber \\
 & = &
  \frac{Q(r_{min})}{\sqrt{a}}\int_{\delta}^{\bar{\epsilon}}d\epsilon\frac{1}{\sqrt{\left(\epsilon+\frac{b}{2a}\right)^{2}-\frac{b^{2}}{4a^{2}}+\frac{c}{a}}}\label{eq:integral}
\end{eqnarray}
where $Q, H$ is same definition as
sec.\ref{sec:Bulk_Background}, and
\begin{eqnarray*}
a & = & \left\{
         \frac{1}{2}\left(Q^{2}\right)^{\prime\prime}H+\left(Q^{2}\right)^{\prime}H^{\prime}\right\}
_{r=r_{m}}\ ,\\
b & = & \left\{
         \left(Q^{2}\right)^{\prime}H-\left(Q^{2}\right)^{\prime}H^{\prime}\delta\right\}
_{r=r_{m}}\ ,\\
c & = & \left\{
-\left(Q^{2}\right)^{\prime}H\delta-\frac{1}{2}\left(Q^{2}\right)^{\prime\prime}H\delta^{2}\right\}
_{r=r_{m}}\ .
\end{eqnarray*}
We can calculate integral (\ref{eq:integral}) by changing integration
variable as $\tilde{\epsilon}=\epsilon+\frac{b}{2a}$, and then result is

\begin{eqnarray}
I(\bar{\epsilon},\delta) \! = \! \frac{Q(r_{min})}{\sqrt{a}}\left\{ \log\left(\bar{\epsilon}+\frac{b}{2a}+\sqrt{\bar{\epsilon}^{2}+\frac{b}{a}\bar{\epsilon}+\frac{c}{a}}\right)
-\log\left(\delta+\frac{b}{2a}+\sqrt{\delta^{2}+\frac{b}{a}\delta+\frac{c}{a}}\right)\right\}\
.\nonumber \\  \label{eq:L_near_rm}
\end{eqnarray}
In the limit of $\delta \to 0$, this expression becomes
\begin{eqnarray}
I(\bar{\epsilon},0)
& = &
\frac{Q(r_{min})}{\sqrt{a}}
\left\{ \log\left(\bar{\epsilon}+p+\sqrt{\bar{\epsilon}^{2}+2p\bar{\epsilon}}\right)-\log\left(p\right)\right\}
\label{eq:L_delta_0}
\end{eqnarray}
and
\begin{eqnarray}
p\equiv \frac{b}{2a} & = & \left.\frac{\left(Q^{2}\right)^{\prime}H}{\left(Q^{2}\right)^{\prime\prime}H+2\left(Q^{2}\right)^{\prime}H^{\prime}}\right|_{r=r_{m}}\nonumber \\
 & = &
\left.\frac{1}{\left(Q^{2}\right)^{\prime\prime}/\left(Q^{2}\right)^{\prime}+2H^{\prime}/H}\right|_{r=r_{m}}\
.\label{eq:def-p}
\end{eqnarray}
If the prefactor of (\ref{eq:L_delta_0}) is finite, we can estimate the
divergency of this expression by estimating $p=b/(2a)$.
There might be 
several ways setting this variable to zero. 
Our
 non-SUSY model achieves this due to $(Q^2)^\prime=0$, whereas Witten's
 D$4$ model accomplish this due to $H/H^\prime=0$. We will see these in the
 rest of this appendix.

First, we see about our non-SUSY model. Since our model has the
same metric in $t$ and $x$ direction, then
$(Q^2)^\prime$ means $(G_{ii}^2)^\prime=2G_{ii}G_{ii}^\prime$ and this is
zero at $r=r_m$ because of $G_{ii}^\prime(r_m)=0$.
We can easily confirm that the only divergent ingredient is $(Q^2)^\prime$ and
others ($(Q^2)^{\prime \prime},H,H^\prime$) are finite at $r_m$,
therefore in (\ref{eq:L_delta_0}) the second term is logarithmic
divergent and its coefficient is finite.

Second, Witten's model has the metric displayed below,
\[
|G_{tt}|=G_{ii}=\left(\frac{U}{R}\right)^{3/2},\ \ \ G_{UU}=\left(\frac{R}{U}\right)^{3/2}\frac{1}{f(U)},\ \ \ f(U)=1-\left(\frac{U_{KK}}{U}\right)^{3}.\]
In this model, there is the  only divergent ingredient,
\begin{eqnarray*}
H & = & G_{ii}/G_{UU}\\
 & = & \left(\frac{U}{R}\right)^{3}f(U)\\
 & = & \frac{U^{3}-U_{KK}^{3}}{R^{3}}\ .
\end{eqnarray*}
Then the second term is logarithmic
divergent and its coefficient is finite in (\ref{eq:L_delta_0}) again,
although the reason of divergence is not the same with the case of our non-SUSY model.
In this case, the cause of the divergence is reduced to $H(U_{KK})=0$ or equally 
to $G_{UU}(U_{KK})=+\infty$. Since it means that
 $G_{ii}(U_{KK})/G_{UU}(U_{KK})=0$, the measure of $i$ direction is
 infinitely small compared to the one of $U$ direction. 

\def\theequation{B.\arabic{equation}}
\setcounter{equation}{0}

\section{More on Wall} \label{more-on-wall}

\vspace{.3cm}
In this appendix, we demonstrate that the infrared wall which suggested
in sec.\ref{sec:Bulk_Background} for classical open string configuration
also arise in other classical configurations. The analysis is performed
for two sorts of classical D-brane configuration, that is, embedded D7
brane and D5 brane.

\subsection{Wall for D7 brane} 

Here, we consider the role of the wall in the D7 embedding. The embedding procedure is
briefly reviewed. 
The world-vomule of D7-brane is set by rewriting the extra six
dimensional part of (\ref{finite-c-sol}) as
\bea
\frac{R^2}{r^2}dr^2+R^2d\Omega_5^2&=&\frac{R^2}{r^2}\left(dr^2+r^2d\Omega_5^2\right)=\frac{R^2}{r^2}\left(\sum_{i=4}^9(dX^{i})^2\right)\\
&=&\frac{R^2}{r^2}\left(d\rho^2+\rho^2d\Omega_3^2+\sum_{i=8}^9(dX^{i})^2\right)\ .
\eea
The worldvolume coordinates $\xi^{M} (M=0\sim 7)$ of D7-brane are taken as $x^{\mu}$ ($\mu=0,\cdots,3$)
and $(\rho,S^3)$, then its
induced metric is expressed as 
\beq
ds_8^2=e^{\Phi/2}\left(\frac{r^2}{R^2}A^2(r)\eta_{\mu\nu}dx^{\mu}dx^{\nu}+\frac{R^2}{r^2}\left((1+w'(\rho)^2)d\rho^2+\rho^2d\Omega_3^2\right)\right)\ ,
\eeq
where $r^2=\rho^2+w(\rho)^2$ and $w'=\partial_{\rho}w$. Here the D7-brane is embedded under the
ansatz,
\beq
(X^8)^2+(X^9)^2=w(\rho)^2\ .
\eeq
We can set the solution of $w(\rho)$ as $(X^8,X^9)=(w(\rho),0)$ since the
background is symmetric under the rotation on $X^8$-$X^9$ plane.
Then the DBI action is expressed as 
\beq
S_{D7}=-\tau_7\int d^8\xi e^{\Phi}A(r)^4\rho^3\sqrt{1+w'(\rho)^2}\, ,
\eeq
where $\tau_7$ denotes the tention of D$7$-brane.

\vspace{.3cm}
The equation of motion for $w(\rho)$ has been solved as
\beq
w(\rho)=m_q+\frac{C}{\rho^2}+\cdots,\quad \rho\to\infty ,
\eeq
where $m_q$ and $C$ denotes the current quark mass and 
the chiral condensate $\langle \bar{\psi}\psi\rangle$ respectively as known from the 
dictionary of the AdS/CFT correspondence. In the non-supersymmetric case, 
we notice that the chiral
symmetry is spontaneously broken \cite{GY}, namely $C>0$ for $m_q=0$.

In the supersymmetric case, we give a comment on the D7 brane embedding. In this case,
the Chern-Simons term is added to the action in our model due to the 
existence of non-trivial zero-form field, namely
the axion. In this case, the lowest energy embedding is given by \cite{GY}\footnote
{Since an argument for this solution has recently been 
given in \cite{Sin}, we give the details of the derivation of this result in the appendix \ref{ap:D7_emmbedding}
. In the appendix, we can see that the results given in \cite{GY} would not be altered.}
\beq
  w=m_q={\rm constant}\, .
\eeq

\vspace{.3cm}
Here, however, we concentrate on another quantity
$w(0)$ of the solution for various quark mass $m_q$. The value of $w(0)$
represents the lowest value of $r$ on the embedded D7 brane. In a sence, therefore,
$w(0)$ corresponds to the infrared cutoff, then it would determine the mass scale of the theory,
for example the meson mass. In the supersymmetric case, the meson mass is 
actually given as \cite{KMMW}
\footnote{In this case, $q=0$, but the situation is similar to the case of $q>0$, since $q$
does not provide any infrared cutoff as seen above. The explicit $q$-dependence is examined
in \cite{BGN}.}
\beq
  M={2m_q\over R^2}\sqrt{(n+1)(n+2)}
\eeq
for radial ($n$) exited spectra. This is interpreted as
the reflection of a finite infrared cutoff scale
$m_q=w(0)$. Then the mass spectra would disappear in the limit of $m_q\to 0$ in this case.
The similar phenomenon is seen for the glueballs as shown below.

\begin{figure}[htbp]
\vspace{.3cm}
\begin{center}
\includegraphics[width=7cm]{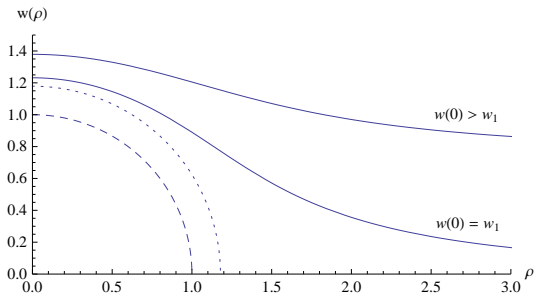}
 \caption{{\small The typical solutions of D7-brane embedding in the non-supersymmetric
 background for $r_0=1$. The broken line represents the singular point at
  $r=r_0$ and the dotted line represents 
  $r=r_m$. The solution of $w(0)=w_1$ corresponds to massless quark at UV
  boundary. We can see that the relation $w(0)>r_m$ is satisfied for any quark mass.  
 }}
\label{D7-sol}
\end{center}
\end{figure}

\vspace{.3cm}
On the other hand, in the non-supersymmetric case,
we can assure that the finite value of $w(0)(>r_m)$ is seen even if
$m_q=0$ due to the positive chiral condensate.  
In the Fig. \ref{D7-sol}, typical examples of the solution
are shown. As seen from this figure,
we find the limitting value of $w(0)(\geq w_1)$ as $w_1=1.2318 r_0$ for $R=1$.
This is the reflection of the wall, which
blocks the
classical D7 brane configuration, which is restricted to the region $r\geq w_1$. 
The meson with small
spin is obtained by the quantum fluctuations of the fields on the D7 brane, and they are also
blocked by this infrared wall for any quark mass $m_q$. 
Since the nearest point of D7 brane is cut at $w_1$ outside
of $r_m$ even if $m_q$ is zero, the meson mass obtained from the D7 brane fluctuation get a
finite mass due to this mass gap.


\subsection{Wall for D5 brane; Baryon vertex} \label{p:baryon}
Furthermore, we can see that the D5 brane, which is introduced as the vertex of baryons \cite{wit}, is also
blocked by the wall. 
The D5-brane wraps $S^5$ in the bulk $M^5\times S^5$, and its action is given as \cite{Callan:1998iq}
\beq
S_{D5}=-\tau_5\int d^6\xi e^{-\Phi}\sqrt{-\det(G_{MN}\partial_{a}X^{M}\partial_{b}X^{N}+\tilde{F}_{ab})}+\tau_5\int A_{(1)}\wedge
G_{(5)}\ ,
\eeq
where $\tau_5$ denotes D5-brane tention,
$\tilde{F}_{ab}=2\pi\alpha'F_{ab}$, $A_{(1)}$ is the U(1) gauge field on D5-brane, and $G_{(5)}$
represents the 5-form self-dual field strength of stacked D3-branes.

After performing the Legendre transformation with respect to the gauge field
$A_{t}$ to eliminate itself, the action is rewritten as \cite{Callan:1998iq}
\beq
U=\frac{N}{3\pi^2\alpha'}\int
d\theta\sqrt{|G_{tt}|G_{rr}(r^{\prime 2}+r^2)}\sqrt{D(\theta)^2+\sin^8\theta}\ ,\label{new-d5-action}
\eeq 
where $N/(3\pi^2\alpha')=T_5\Omega_4 R^4$. The equation of motion for $r(\theta)$ of the
lagrangian (\ref{new-d5-action}) is obtained as 
\bea\label{d5-eom}
\partial_{\theta}\left(\frac{\sqrt{U(r)}\sqrt{V(\nu,\theta)}r'}{\sqrt{r^{\prime
2}+r^2}}\right)-\frac{\frac{1}{2}\frac{\partial U(r)}{\partial
r}(r^{\prime 2}+r^2)+U(r)r}{\sqrt{U(r)}\sqrt{r^{\prime 2}+r^2}}=0\ ,
\eea
where we define $U(r)\equiv|G_{tt}|G_{rr}$. In general, the lowest
energy configuration is the point-like solution for $r$-direction given 
by $r(\theta)=r_b$, where $r_b$ is a constant.
Assuming that $U(r_b)\neq 0$ and $r_b\neq 0$, the value of $r_b$ is obtained 
as follows:
\beq
U'(r_b)r_b+2U(r_b)=0.
\eeq
In the non-SUSY background (\ref{finite-c-sol}), we find
\beq
r_b=\left(\sqrt{6}+\sqrt{5}\right)^{1/4}r_0\approx 1.47r_0\ .
\eeq
Then the classical configuration of the vertex is also trapped at the point 
above $r_m(=1.18r_0)$.


\def\theequation{C.\arabic{equation}}
\setcounter{equation}{0}

\section{D7 brane embedding and eight form} \label{ap:D7_emmbedding}



The supersymmetric solutions for the axion $\chi$ and dilaton $\Phi$ 
used here are obtained under the ansatz 
\cite{KS2,LT},
\beq
\chi=-e^{-\Phi}+\chi_0 \ ,
\label{super2}
\eeq
which is necessary to obtain supersymmetric solutions. 
The metric is expressed as
\beq
  ds^2_{10}=e^{\Phi/2}
\left\{
{r^2 \over R^2}\left(-dt^2+(dx^i)^2\right)+
\frac{R^2}{r^2}\left( d\eta^2+\eta^2 d\Omega_3^2
+(dX^8)^2+(dX^9)^2\right) \right\} \ . 
\label{finite-c-sol-2}
\eeq
In this coordinate, the solution is obtained as $\Phi=\Phi(r)$ and $r^2=\eta^2+(X^8)^2+(X^9)^2$, so the nine form
field strength dual to the $\chi$ is given as
$ F_{(9)}= *d\chi$ 
and
$$
 F_{(9)}= g^{\eta\eta}\sqrt{-g} (\partial_{\eta}\chi~dt\wedge d\vec{x}
\wedge d\Omega_3\wedge dX^8\wedge dX^9+
\partial_8 \chi~{dt\wedge d\vec{x}\wedge d\eta\wedge d\Omega_3\wedge dX^9}$$
\beq\label{F-10}
-\partial_9 \chi~{dt\wedge d\vec{x}\wedge d\eta\wedge d\Omega_3\wedge dX^8})\, 
\eeq
\beq
 \equiv g^{\eta\eta}\sqrt{-g} (\partial_{\eta}\chi~d_{\hat{\eta}}+
\partial_8 \chi~d_{\hat{X}^8}-\partial_9 \chi~d_{\hat{X}^9})
\eeq
where we defined as $\partial_8={\partial\over \partial X^8}$ 
and the outer product as
\beq
  d_{\hat{\eta}}\equiv dt\wedge d\vec{x}
\wedge d\Omega_3\wedge dX^8\wedge dX^9
\eeq
etc.
Here we notice $ g^{\eta\eta}= g^{yy}=(r/R)^2$ and $\epsilon_{t\vec{x}\eta\Omega_3 89}=1$. 

\vspace{.2cm}
Introducing the eight form in the form
\beq
 C_{(8)}=f_8(\eta,X^8,X^9)d_{\hat{\eta}\hat{X}^9}+
 f_9(\eta,X^8,X^9)d_{\hat{\eta}\hat{X}^8}+g_{\eta}(\eta,X^8,X^9)d_{\hat{X}^8\hat{X}^9}\, ,
\eeq
the nine form field strength is also obtained by $F_{(9)}=d C_{(8)}$ as
\beq\label{F-11}
 F_{(9)}=\partial_{\eta}f_8 d_{\hat{X}^9}+\partial_9f_8 d_{\hat{\eta}}
 +\partial_{\eta}f_9 d_{\hat{X}^8}-\partial_8f_9 d_{\hat{\eta}}
+\partial_{8}g_{\eta} d_{\hat{X}^9}+\partial_9g_{\eta} d_{\hat{X}^8}\, .
\eeq
Comparing Eqs.(\ref{F-10}) and (\ref{F-11}), we find
\bea 
 \partial_{9} f_8-\partial_{8} f_9&=&g^{\eta\eta}\sqrt{-g}\partial_{\eta}\chi\, ,  \label{c8-1} \\
 \partial_{9} g_{\eta}+\partial_{\eta} f_9&=&g^{\eta\eta}\sqrt{-g}\partial_{8}\chi\, , \label{c8-2} \\
 \partial_{8} g_{\eta}+\partial_{\eta}f_8&=&-g^{\eta\eta}\sqrt{-g}\partial_{9}\chi\, . \label{c8-3} 
\eea
Noticing 
\beq
  g^{\eta\eta}\sqrt{-g}=\sqrt{\epsilon_3}\eta^3 e^{2\Phi}\, ,
\eeq
where $\epsilon_3$ denotes the metric of $S^3$ part, and using 
(\ref{super}) and $e^{\Phi}=1+q/r^4$ we obtain
\beq
 g^{\eta\eta}\sqrt{-g}\partial_{8}\chi=\sqrt{\epsilon_3}\eta^3 e^{2\Phi}\partial_{8}\chi
        =-4\sqrt{\epsilon_3}\eta^3{qX^8\over r^6}\, .
\eeq
Then $f_9$ is obtained by solving (\ref{c8-2})
\beq \label{c8-9} 
  f_9=\sqrt{\epsilon_3}qX^8\left({1\over r^2}+{\eta^2\over r^4}\right)+C_9\, ,
\eeq 
where 
\beq
  C_9=-\int d\eta~\partial_{9}g_{\eta}\, .
\eeq
In this indefinite integration with respect to $\eta$, we can add an arbitrary function of $X^8$ and $X^9$. We consider that it
is included in $C_9$ here. Similarly, we obtain $f_8$ from (\ref{c8-3}) as
\beq \label{c8-8} 
  f_8=-\sqrt{\epsilon_3}qX^9\left({1\over r^2}+{\eta^2\over r^4}\right)+C_8\, ,
\eeq 
\beq
  C_8=-\int d\eta~\partial_{8}g_{\eta}\, .
\eeq
Then the remaining Eq.(\ref{c8-1}) is rewritten by using (\ref{c8-8}) and (\ref{c8-9}) as
\beq
 \partial_9C_8-\partial_8C_9=0\, .
\eeq
While this gives a constraint on the arbitrary functions of $X^8$ and $X^9$ added to
$C_8$ and $C_9$, it is however independent of $g_{\eta}$ since we can see
\beq
 \partial_9C_8-\partial_8C_9=-\partial_9\int d\eta~\partial_{8}g_{\eta}
+\partial_8\int d\eta~\partial_{9}g_{\eta}=0\, ,
\eeq
where we ignored the added arbitrary functions of $X^8$ and $X^9$. In other words,
we cannot get any constraint for $g_{\eta}$, which is the main part of the
Chern-Simons term of the D7 brane action. Here, it is determined as follows.

\vspace{.2cm}
The pull backed eight form fields are written for our world volume of D7 brane as 
\beq
  C_{[8]}\equiv\tilde{g}_{(8)}d_{\hat{X}^8\hat{X}^9}
         =(g_{\eta}-f_8\dot{X}^8-f_9\dot{X}^9)d_{\hat{X}^8\hat{X}^9}\, ,
\eeq
where the dot denotes the derivative with respect to $\eta$, for example
$\dot{X}^8=\partial_{\eta}X^8$, and
the action of the D7 brane is given as
\beq
  S_{D7}=-\tau_7\int d\xi^8\left(e^{-\Phi}\sqrt{-{\cal G}}+\tilde{g}_{(8)}\right)\, .
\eeq
Here ${\cal G}$ denotes the determinant of the induced metric of D7 brane,
which is taken as
\bea\label{induced D7}
  ds^2_{8}&=&e^{\Phi/2}
\left\{
{r^2 \over R^2}\left(-dt^2+(dx^i)^2\right)+
\frac{R^2}{r^2}\left( (1+(\partial_{\eta}X^8(\eta ))^2 
 +(\partial_{\eta}X^9(\eta ))^2 ) d\eta^2 \right. \right. \nonumber \\ 
     && \qquad ~~~~~+\left. \left. \eta^2 d\Omega_3^2
+(dX^8)^2+(dX^9)^2\right) \right\} \ ,
\eea
where we notice $r^2=\eta^2+{X_8}^2+{X_9}^2$.
Then ${\cal G}$ is given as,
\beq
 \sqrt{-{\cal G}}=\sqrt{\epsilon_3}\eta^3 e^{2\Phi}
\sqrt{1+(\partial_{\eta}X^8(\eta))^2+(\partial_{\eta}X^9(\eta))^2}\, .
\eeq

\vspace{.3cm}
In the above action, the Chern-Simons part $\tilde{g}_{(8)}$ is given as
\bea
 \tilde{g}_{(8)}&=&{g}_{\eta}-f_8\dot{X}^8-f_9\dot{X}^9\, \label{g8-1}\\
 &=& {g}_{\eta}+q\left({1\over r^2}+{\eta^2\over r^4}\right)
     \left(X^9\dot{X}^8-X^8\dot{X}^9\right)
    -C_8\dot{X}^8-C_9\dot{X}^9\, ,  \label{g8-2}
\eea
where however ${g}_{\eta}$, $C_8$, and $C_9$ are not given explicitely
since they are not determined. These undetermined functions would be given
by appropriate boundary conditions of the system.

\vspace{.3cm}
Our purpose is to find an embedding solution of the D7 brane. We try it by
an ansatz to obtain a simple solution. Consider the parametrization,
\beq
  X^8=w(\eta)\cos\theta\, , \quad X^9=w(\eta)\sin\theta\, ,
\eeq 
where $\theta$ is a constant and independent of $\eta$. The embedding is
given by the profile of $w(\eta)$. In general, the solution has the following asymptotic form at ${\eta\to\infty}$ 
\beq
  w(\eta)=m_q+{c\over\eta^2}+\cdots
\eeq 
where $m_q$ and $c$ represent the current quark mass and the chiral order
parameter, $\langle\bar{\psi}\psi\rangle$ with the quark filed $\psi$.
Further simplification is done by setting $\theta=0$ or $\theta=\pi/2$.
Here we take $\theta=0$, then
\beq\label{g8-3}
 \tilde{g}_{(8)}={g}_{\eta}-C_8\dot{w}\, .  
\eeq
This is rewritten by the partial integration with respect to $\eta$ as
\beq\label{g8-4}
 \tilde{g}_{(8)}={g}_{\eta}+w\dot{C}_8={g}_{\eta}-w\partial_w{g}_{\eta}
\equiv \Omega_3\bar{g}_{(8)}\, .  
\eeq
where $\Omega_3$ is the volume of $S^3$ and 
we used $C_8=-\int d\eta \partial_w g_{\eta}$. Then the effective
D7 brane action is written as
\beq\label{d7action-2}
  S_{D7}=-\tau_7\Omega_3\int dx^4 d\eta\left(\eta^3 e^{\Phi}
\sqrt{1+\dot{w}^2}+\bar{g}_{(8)}\right)\, .
\eeq

For simplicity, we assume as $g_{\eta}=g_{\eta}(\eta,w(\eta))$, then
$\bar{g}_{(8)}=\bar{g}_{(8)}(\eta,w(\eta))$. Form (\ref{d7action-2}),
the equation of motion of $w$ is obtained as
\beq\label{w-eq}
  \eta^3\partial_{w}\left(e^{\Phi}\right)\sqrt{1+(\dot{w})^2}
  -\partial_{\eta}\left(\eta^3e^{\Phi}{\dot{w}\over\sqrt{1+(\dot{w})^2}}\right)
  =-\partial_{w}\bar{g}_{(8)}\, .
\eeq
Here we demand that there should exist a supersymmetric embedding, namely
a constant $w$ (or $\dot{w}=0$) is the solution. This implies our previous
Chern-Simon term \cite{GY}
\beq
  \bar{g}_{(8)}=-\eta^3 e^{\Phi}\, ,
\eeq 
where we have neglected $w$-independent terms. We should say that this
result is not unique of course. There are other setting of $\bar{g}_{(8)}$
which does not allow the supersymmetric solution of $w$. In such cases, however,
we should find some dynamical origin of supersymmetry breaking in the dual theory.
It may be an interesting problem, but it is opened here.



\newpage
\end{document}